\documentclass[useAMS,usenatbib]{mn2e}

\usepackage{times}
\usepackage{multirow}
\usepackage{amsmath}
\usepackage[dvips]{graphicx}

\def\pcm3{{\rm\thinspace cm^{-3}}}

\def\contcaption{\@conttrue\SFB@caption\@captype}

\def\n_h{{\rm n_{H}}}

\def\NH1{{$N_{\rm HI}~$}}

\def\ga{{\rm\thinspace gauss}}

\def\approxlt{\mathrel{\hbox{\rlap{\lower .5ex \hbox {$\sim$}}
        \raise .15 ex \hbox{$<$}}}}
\def\approxgt{\mathrel{\hbox{\rlap{\lower .5ex \hbox {$\sim$}}
        \raise .15 ex \hbox{$>$}}}}

\def\la{\mathrel{\hbox{\rlap{\hbox{\lower4pt\hbox{$\sim$}}}\hbox{$<$}}}}
\def\ga{\mathrel{\hbox{\rlap{\hbox{\lower4pt\hbox{$\sim$}}}\hbox{$>$}}}}
\newbox\grsign \setbox\grsign=\hbox{$>$} \newdimen\grdimen
\grdimen=\ht\grsign
\newbox\simlessbox \newbox\simgreatbox \newbox\simpropbox
\setbox\simgreatbox=\hbox{\raise.5ex\hbox{$>$}\llap
     {\lower.5ex\hbox{$\sim$}}}\ht1=\grdimen\dp1=0pt
\setbox\simlessbox=\hbox{\raise.5ex\hbox{$<$}\llap
     {\lower.5ex\hbox{$\sim$}}}\ht2=\grdimen\dp2=0pt
\setbox\simpropbox=\hbox{\raise.5ex\hbox{$\propto$}\llap
     {\lower.5ex\hbox{$\sim$}}}\ht2=\grdimen\dp2=0pt
\def\simgreat{\mathrel{\copy\simgreatbox}}
\def\simless{\mathrel{\copy\simlessbox}}

\title[SMC red giant stars - kinematics]{Red Giants in the Small Magellanic Cloud. I. Disk and Tidal Stream Kinematics}

\author[Dobbie et al.]{P.~D.~Dobbie$^{1}$\thanks{E-mail: paul.dobbie@utas.edu.au}, A.~A.~Cole$^{1}$, A.~Subramaniam$^{2}$, 
S. Keller$^{3}$ \newauthor  \\
$^{1}$School of Physical Sciences, University of Tasmania, Hobart, TAS, 7001, Australia\\
$^{2}$Indian Institute of Astrophysics, Bengaluru 560034, India \\
$^{3}$Research School of Astronomy and Astrophysics, Australian National University, Canberra, Australia \\
}

\begin{document}

\date{Accepted; Received; in original form}

\pagerange{\pageref{firstpage}--\pageref{lastpage}} \pubyear{2009}

\maketitle

\label{firstpage}

\begin{abstract}

We present results from an extensive spectroscopic survey of field stars in the Small Magellanic Cloud (SMC). 3037 sources, predominantly first-ascent red giants, spread across roughly 37.5 deg$^2$, are analysed. The line of sight velocity field is dominated by the projection of the orbital motion of the SMC around the LMC/Milky Way. The residuals are inconsistent with both a non-rotating spheroid and a nearly face on disk system. The current sample and previous stellar and HI kinematics can be reconciled by rotating disk models with line of nodes position angle $\Theta$ $\approx$ 120$-$130$^{\circ}$, moderate inclination (25$-$70$^{\circ}$), and rotation curves rising at 20$-$40 km~s$^{-1}$~kpc$^{-1}$. The metal-poor stars exhibit a lower velocity gradient and higher velocity dispersion than the metal-rich stars. If our interpretation of the velocity patterns as bulk rotation is appropriate, then some revision to simulations of the SMC orbit is required since these are generally tuned to the SMC disk line-of-nodes lying in a NE-SW direction.  Residuals show strong spatial structure indicative of non-circular motions that increase in importance with increasing distance from the SMC centre. Kinematic substructure in the north-west part of our survey area is associated with the tidal tail or Counter-Bridge predicted by simulations. Lower line-of-sight velocities towards the Wing and the larger velocities just beyond the SW end of the SMC Bar are probably associated with stellar components of the Magellanic Bridge and Counter-Bridge, respectively.  Our results reinforce the notion that the intermediate-age stellar population of the SMC is subject to substantial stripping by external forces.

\end{abstract}


\section{Introduction}

One of the principal goals of contemporary astrophysics is to develop a more complete understanding of galaxy formation and evolution. 
While the prevailing theoretical framework, Lambda-Cold Dark Matter \citep[$\Lambda$-CDM, e.g.][]{peebles03}, is rather successful in 
replicating the large scale structures observed in the Universe \citep[e.g.][]{springel05,coles05} it suffers significant shortcomings 
at explaining smaller scale phenomena where density perturbations depart strongly from the linear regime and the role of baryon physics 
becomes substantial \citep[e.g.][]{kroupa10,famaey13}.
In $\Lambda$-CDM cosmology, galaxy formation is a hierarchical process in which the larger structures grow through the aggregation of 
small dark matter haloes and baryons. As the dissipative gas cools, it collapses to densities sufficient for star formation to occur 
\citep{tegmark97}. The ongoing accretion of gas with higher specific angular momentum promotes an inside-out development of galactic 
disks and the gradual migration in time of star formation activity to larger galacto-centric radii. This is in accord with observations
of negative radial chemical abundance gradients in the disks of local spiral galaxies and the formation of stars with comparatively low
metallicities in their outer disk regions in the present epoch \citep[e.g.][]{wang11}. 

However, the basic theoretical framework overpredicts the numbers of dwarf galaxies in the Local Volume, including the number that are 
satellites to the Milky Way \citep[][]{klypin99}. While the mass function of galaxies might be expected to be similar in shape to that of 
the dark matter haloes, ie. proportional to $M^{-1.9}$, it is observed to be closer in form to $M^{-1}$ in the low luminosity regime \citep{coles01}.
In addition, observational studies of the rotation curves of dwarf galaxies show them to be slowly rising with increasing galacto-centric 
distance \citep{deblok02}, indicative of a dark matter distribution that is sigificantly flatter than the centrally cusped form predicted by
$\Lambda$-CDM. To account for these disparities, several mechanisms have been invoked that can both regulate the formation of stars in galaxies 
and smooth out the central cusp in their dark matter distributions. These include internal factors such as supernovae feedback \citep[][]{governato10}, 
and external influences such as background UV radiation and tidal and/or ram-pressure 
stripping of potentially star forming gas from a system by galaxy-galaxy interactions \citep[][]{kazantzidis13}. For example, observations of 
the Fornax galaxy cluster indicate that environmental factors regulate the levels of star formation activity in the dwarf members \citep[][]{drinkwater01}.

These regulating mechanisms have also been linked to the apparently discordant outside-in progression of star formation in many low luminosity dwarf 
irregular systems \citep[e.g.][]{zhang12}. Deep imaging studies reveal recent star formation to be concentrated within their central 
regions \citep[e.g. Phoenix, IC\,1613 and NGC\,6822][]{hidalgo09,skillman03,wyder01}, suggesting that accretion of high angular momentum gas is
inhibited. The reduction in turbulent gas pressure in the denser inner 
parts of these galaxies following the supernovae blow out of disk material, is suspected to lead to the inward migration of enriched gas and the contraction 
towards their centers of the star forming disk \citep[e.g.][]{pilkington12, stinson09}. Additionally, tidal interactions may incite bar like 
instabilities in these galaxies that can promote the inwards flow of gas in their disks. 

Despite being somewhat less common than predicted by theory, dwarfs still numerically dominate the galaxy population. As the antecendents of larger galaxies such as 
Messier 31 and the Milky Way, it is vital to understand their architectures and evolution, including the roles of disk rotation and pressure support, their 
dark matter distributions and the regulation of their star formation. The Small Magellanic Cloud (SMC) is the smaller of a pair of comparatively massive ($M$$>$10$
^{9}$M$_{\odot}$) dwarf galaxies close ($D$$\le$60kpc) to the Milky Way. As probable satellites of the Galaxy they are relatively unusual in that they are gas 
rich whereas the majority of dwarf galaxies within 270kpc of the Milky Way and Messier 31 appear to be gas poor \citep[e.g.][]{grcevich09}. However, 
there is substantial evidence that gas is being stripped from the Magellanic Clouds as a consequence of their interactions with the Galaxy and each other. 
For example, they are immersed within an extended body of diffuse HI gas that stretches out many tens of degrees across the sky, forming the Magellanic Stream 
and the Leading Arm \citep[e.g.][]{putman03}.

\begin{figure}
\includegraphics[angle=0, width=\linewidth]{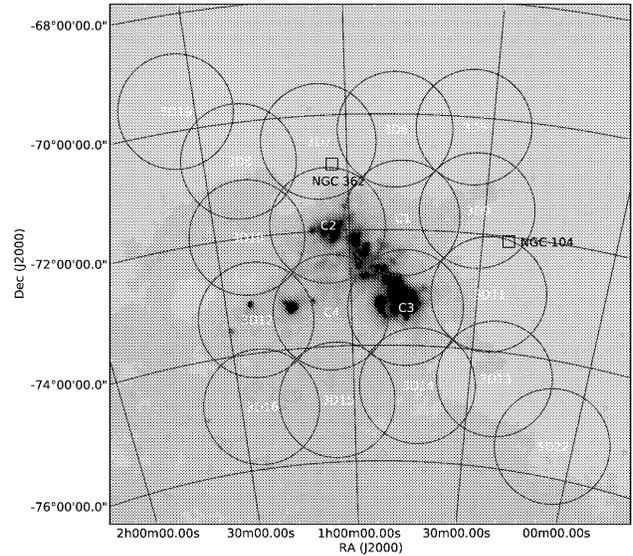}
\caption{A 9$^{\circ}$$\times$9$^{\circ}$ image of the sky centered on the SMC \citep[100$\mu$m IRIS data, ][]{mdl05}. The 
circles, which each corresponds to an AAT + 2dF/AAOmega pointing, highlight the areas included in our photometric and 
spectroscopic survey. Note that at least two distinct fibre configurations were observed for each of the four central fields. Two 
Galactic globular clusters within our survey area are also highlighted (open squares).}
\label{smcmap}
\end{figure}

In HI observations the SMC displays a ``frothy'' appearance, attributed to a large number of recent supernova explosions, and a substantial velocity gradient
along a position-angle (PA)$\approx$60$^{\circ}$, which has been associated with the systemic rotation of a cold disk of gas \citep{stanimirovic04}.
The young and the 
intermediate/old stellar populations of the Cloud display quite distinctive morphologies. The former have an irregular distribution and it has been inferred from 
observations of Cepheids that the main body of the SMC, where much of this stellar population resides, corresponds to a bar structure that is being viewed virtually
end on \citep{caldwell85}. The south-west end of the main body is believed to be slightly more distant than the north-east although this latter region appears to 
consist of two distinct kinematic structures lying at different distances \citep{hatzidimitriou93}. The old/intermediate stellar population appears to be much more
evenly distributed \citep{zaritsky00} and recent observations suggest it extends many degrees from the center of the Cloud \citep{nidever11}. Moreover, its kinematical 
properties appear to be consistent with those of a pressure supported spheroid \citep{harris06}. The contrast between the distributions of the young and the 
older populations have led to suggestions that the former is the outcome of a recent gas infall event \citep{zaritsky00,zaritsky04}, while \cite{subramanian12} have 
proposed that a dwarf-dwarf merger occured between 2--5Gyr ago. 

\begin{figure}
\includegraphics[width=\linewidth]{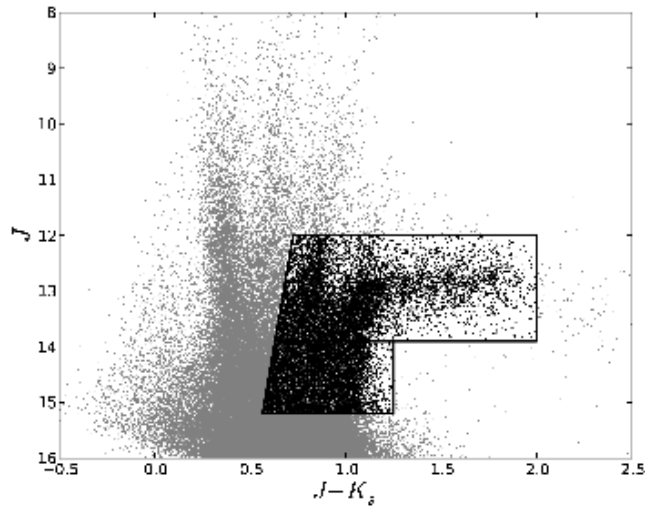}
\caption{A 2MASS near-IR colour-magnitude plot of point sources in an area of roughly 37.5 deg$^{2}$ towards the SMC. Sources selected for potential 
spectroscopic follow-up were drawn from the highlighted region.}
\label{CMD}
\end{figure}

Several key observational properties of the SMC are qualitatively reproduced by N-body and chemo-dynamical modelling in which it interacts with
the Large Magellanic Cloud (LMC) and the Galaxy, including the velocity field of the Magellanic Stream, the Magellanic Bridge structure towards
the LMC, the kinematics and the distribution of the intermediate/old stellar population, the large line of sight depth of the Cloud and the 
age-metallicity relation \citep[][]{murai80,gardiner96,yoshizawa03, bekki09}. Realistic simulations are particularly important for reconstructing
the interaction history of the Magellanic Clouds that can lead to a deeper understanding of the impact of tidal and ram pressure forces on the 
structure and the evolution of the SMC and dwarf galaxies in general. In addition, these computations, through accurately reproducing the properties 
of the Magellanic Stream, can afford further insight on the distribution of the dark matter halo of the Galaxy \citep{haghi06}. 

\begin{table*}
\begin{minipage}{130mm}
\begin{center}
\caption{Details of the field configurations used for obtaining the spectroscopic follow-up observations 
of candidate SMC red-giant stars.}
\label{AAOmega}
\begin{tabular}{lccccccc}
\hline
Name & RA & Dec & $t_{\rm exp}$ & $n_{\rm exp}$ & Obs. date  & Plate ID & Seeing \\ 
      & hh:mm:ss & ${^\circ}$:$'$:$''$ & & & & & (arcsec)\\
\hline

C1FA   & 00:48:14 & -71:47:59 & 1200 & 3 & 2011/10/18 & 0 &   2\\
C1FB   & 00:48:14 & -71:47:55 & 1200 & 3 & 2011/10/20 & 1 &   2\\
C2FA   & 01:04:47 & -71:54:19 & 1200 & 3 & 2011/10/18 & 1 &   1.5\\
C2FB   & 01:04:47 & -71:54:25 & 1200 & 4 & 2011/10/20 & 0 &   2\\
C3FA   & 00:46:59 & -73:20:44 & 1200 & 3 & 2011/10/18 & 0 &   1\\
C3FB   & 00:47:00 & -73:20:46 & 1200 & 4 & 2011/10/20 & 1 &   1.5\\
C3FC   & 00:46:59 & -73:20:44 & 1200 & 3 & 2011/10/21 & 1 &   1.5\\
C4FA   & 01:05:00 & -73:24:27 & 1200 & 2 & 2011/10/18 & 1 &   1.5\\
C4FB   & 01:04:59 & -73:24:32 & 1200 & 3 & 2011/10/20 & 0 &   2\\
3D05FA & 00:33:53 & -70:10:18 & 1200 & 3 & 2011/10/21 & 0 &   1.5 \\
3D06FA & 00:49:58 & -70:16:05 & 1200 & 4 & 2011/10/20 & 0 &   1.5 \\
3D07FA & 01:05:50 & -70:26:53 & 1200 & 3 & 2011/10/19 & 1 &   1.5\\
3D08FA & 01:22:40 & -70:40:10 & 1200 & 3 & 2011/10/19 & 1 &   1.5\\
3D09FA & 00:31:46 & -71:36:12 & 1200 & 3 & 2011/10/19 & 0 &   2\\
3D10FA & 01:22:57 & -71:59:36 & 1200 & 3 & 2011/10/19 & 0 &   1.5\\
3D11FA & 00:27:18 & -73:01:35 & 1200 & 3 & 2011/10/18 & 0 &   2.5\\
3D12FA & 01:23:15 & -73:25:34 & 1200 & 3 & 2011/10/21 & 0 &   1.5\\
3D13FA & 00:23:34 & -74:28:10 & 1200 & 3 & 2011/10/19 & 0 &   2\\
3D14FA & 00:43:22 & -74:41:34 & 1200 & 3 & 2011/10/18 & 1 &   2\\
3D15FA & 01:04:36 & -74:55:41 & 1200 & 3 & 2011/10/21 & 1 &   1.5\\
3D16FA & 01:24:45 & -74:55:33 & 1200 & 3 & 2011/10/21 & 1 &   1.5\\
5D19   & 01:33:49 & -69:39:53 & 1200 & 3 & 2011/10/21 & 0 &   1.5\\
5D22   & 00:05:35 & -75:27:40 & 1200 & 3 & 2011/10/19 & 1 &   2\\

\hline
\label{temps}
\end{tabular}
\end{center}
\end{minipage}
\end{table*}

As the initial conditions of these simulations are typically determined by integrating the Clouds' orbits backwards in time and through choosing galaxy 
structures and disk orientations that lead to agreement with our understanding of the SMC in the present epoch \citep[e.g.][]{gardiner96}, the limitations 
of current observations and in our knowledge of the orbits contribute to inaccuracies in the inferred evolutionary history. Fortunately much 
improved proper motion determinations are becoming available and are leading to a better definition of the orbits of both the SMC and the LMC 
\citep[e.g.][]{kallivayalil13}. Considering this, it seems timely to re-examine our understanding of the structure and kinematics of the SMC as this could
also help to further refine the simulations. In this vein we have recently performed the most extensive spectroscopic study of the SMC's red-giant population 
to date. Here we present radial velocities for in excess of 3000 stars distributed across an area of roughly 37.5 deg$^{2}$ centered on the Cloud. In 
subsequent sections we outline our initial photometric selection of candidates and our acquisition, reduction and analysis of the spectroscopic follow-up
data. We examine in detail the projected line-of-sight velocity field of the red-giant population to search for evidence of large scale trends. We compare 
our results to prior work on the intermediate/old and the young star populations of the SMC and consider them in the contexts of a disk model and a recent 
tidal interaction.

\section[]{Photometric selection of candidate SMC Red Giant stars}
\label{survey}

An initial selection of candidate SMC red-giants was made from the near-IR photometry of the 2 micron All-Sky Survey 
\cite[2MASS; ][]{skrutskie06} point source catalogue (PSC). A $J$,$J$--$K_{S}$ colour-magnitude diagram was constructed 
for stellar-like sources with photometric uncertainties of less than 0.5 mag. in both $J$ and $K_{S}$,
within an approximately 37.5 deg$^{2}$ region centred on the Cloud (figure~\ref{smcmap}). Sources 
flagged as possible blends, as having photometry contaminated by image artifacts or nearby bright objects and/or as 
lying within the boundaries of catalogued extended sources were excluded. We selected all remaining objects to the red of 
the line defined by $J$ = 26.5 - 20 $\times$ ($J$--$K_{S}$) and blueward of $J$--$K_{S}$=2.0 or $J$--$K_{S}$=1.25 for 12.0$\le$$J$$
<$13.9 and 13.9$\le$$J$$\le$15.2, respectively. These criteria, highlighted in figure~\ref{CMD},
encompass the region of colour-magnitude space spanned by both the red-giant branch (RGB) and asymptotic-giant 
branch (AGB) of the SMC population and led to a preliminary catalogue of 92\,893 sources.

\section{Optical spectroscopy}

\subsection{Observations and data reduction.}

Follow-up optical spectroscopy for a subsample of these stars was acquired during the period 18-21 October 2011, 
with the 2dF/AAOmega instrument and the 3.9m Anglo-Australian Telescope (AAT) located at Siding Spring Observatory, 
Australia. AAOmega is a two arm fibre-fed multi-object optical-spectrograph capable of the simultaneous 
observation of 400 objects distributed over a two degree diameter circular field-of-view \citep{saunders04,sharp06}. 

During this observing campaign, the blue and red arms of the instrument were configured with the 1500V (R$\approx$4000) 
and 1700D (R$\approx$10000) gratings and tuned to central wavelengths of 5350\AA\ and 8670\AA, respectively. This 
provided coverage of the $\lambda$5167, 5172 and 5183\AA\ Mg b and $\lambda$8498, 8542 and 8662\AA\ CaII 
triplet lines. Fortunately, skies were largely clear for much of the run and seeing was generally close
to the Siding Spring Observatory median value. Therefore, during the four nights approximately 
7000 objects were targeted with 23 different field configurations. Details of the pointings, including dates, field centers and
exposure times are reported in Table ~\ref{AAOmega}.

The AAOmega data were reduced using the Australian Astronomical Observatory's {\tt 2dFDR} pipeline which is
described at length by \cite{bailey98} and \cite {sharp10}. In brief, the data were first bias and dark subtracted using
master frames created from exposures taken over the course of the four nights. Fibre-flat exposures of a quartz lamp, obtained 
immediately prior to the science observations of each target, were used to locate the spectra in each CCD frame. The 
fibre-flat field and science spectra were extracted and the latter divided by the former to reduce the impact of 
pixel-to-pixel response variations. Spectra of a CuAr+CuNe+FeAr arc lamp, that were also acquired adjacent in time to the 
science observations, were then used to wavelength calibrate each dataset. Finally, the multiple datasets obtained for the 
targets, typically three per plate configuration (see Table~\ref{AAOmega}), were combined to form the final spectra.

\subsection{Spectroscopic analysis.}
\label{sa}

The spectra from the red-arm of the instrument were first matched to multiplicative combinations of low-order polynomials and normalised 
synthetic spectra drawn from the library of \cite{kirby11}. As these models were calculated in local thermodynamic equilibrium (LTE) and 
do not accurately reproduce the form of the strong, empirical, CaII triplet absorption features, the synthetic Ca lines were augmented 
with Voigt profiles. An iterative approach to fitting was adopted in which a $\chi^{2}$ goodness-of-fit statistic was 
minimised\footnote{http://cars9.uchicago.edu/software/python/lmfit/}, weighting the spectral channels by their inverse variances as determined
by the {\tt 2dFDR} pipeline. Following this step, any points lying more than 5$\sigma$ above or 3$\sigma$ below the model were rejected 
before the data were re-fitted. This procedure was repeated three times and afforded reasonable representations of the datasets of field dwarfs
and RGB stars. A useful additional outcome of this process was a model based estimate for the radial velocity of each target. Subsequently,
to achieve first order normalisation, each spectrum was divided by the low order polynomial component of its corresponding model. As the synthetic
spectral library employed here is not optimised for C-stars, our model representations of the spectra of objects of this nature were of lower 
quality but were sufficient for the purposes here. 

\begin{table}
\begin{minipage}{90mm}
\begin{center}
\caption{Details of the ten RGB radial velocity template stars drawn from the clusters 
NGC\,288, NGC\,362 and Melotte\,66.}
\label{RVs}
\begin{tabular}{lccccccc}
\hline
ID & RA & Dec & v$_{r}$ & References \\ 
      & hh:mm:ss.ss & ${^\circ}$:$'$:$''$ & (/kms$^{-1}$) & \\
\hline

 403 & 00:52:46.25 & -26:37:26.0 & -46.0 & 1,2 \\
 338 & 00:52:52.80 & -26:34:38.8 & -49.2 & 1,2 \\
 344 & 00:52:52.87 & -26:35:20.2 & -49.0 & 1,2 \\
 274 & 00:53:01.13 & -26:36:07.1 & -40.5 & 1,2 \\ \\

 2127 & 01:02:37.64 & -70:50:37.1 & +222.6 & 3,2 \\
 1441 & 01:03:21.73 & -70:48:40.4 & +222.3 & 3,2 \\
 1423 & 01:03:33.01 & -70:49:37.2 & +232.3 & 3,2 \\ \\

4151 & 07:26:12.07 & -47:43:24.7 & +23.0 & 4,5 \\
4266 & 07:26:17.30 & -47:44:00.1 & +21.0 & 4,5 \\
3133 & 07:26:30.53 & -47:41:43.9 & +18.0 & 4,5 \\

\hline
\label{RVs}
\end{tabular}
\end{center}
1. \cite{olszewski84},
2. \cite{shetrone00},
3. \cite{harris82},
4. \cite{hawarden76},
5. \cite{friel02}
\end{minipage}
\end{table}

\begin{table}
\begin{minipage}{90mm}
\begin{center}
\caption{Details of the six SMC C-star radial velocity templates.}
\label{CRVs}
\begin{tabular}{lccccccc}
\hline
ID & RA & Dec & v$_{r}$ & Reference \\ 
      & hh:mm:ss.ss & ${^\circ}$:$'$:$''$ & (/kms$^{-1}$) & \\
\hline

3D05FA 4342  & 00:21:19.5 & -70:54:36 & +139.0 & 1 \\
3D05FA 155   & 00:26:34.3 & -70:14:24 & +158.8 & 1 \\
3D09FA 11102 & 00:29:04.4 & -72:13:17 & +183.6 & 1 \\
3D05FA 3280  & 00:30:47.4 & -70:28:06 & +112.6 & 1 \\
3D05FA 2366  & 00:31:25.3 & -70:23:46 & +132.8 & 1 \\
3D14FA 348   & 01:40:27.0 & -75:41:54 & +163.7 & 1 \\

\hline
\label{CRVs}
\end{tabular}
\end{center}
1. \cite{kunkel97}
\end{minipage}
\end{table}

Next, all the normalised red-arm spectra were cross-correlated with AAOmega data that we obtained for ten RGB objects 
in the clusters NGC\,362, Melotte\,66 and NGC\,288. These stars were observed through a variety of AAOmega's fibres 
and were adopted because reliable radial velocity estimates are available for them in the literature (Table~\ref{RVs}).
The cross-correlation procedure was undertaken with the {\tt IRAF FXCOR} software routine running within a {\tt PyRaf} 
environment. The quoted velocity for each star is a mean of these ten estimates, weighted by their individual errors as 
reported by {\tt FXCOR}. Their associated uncertainties have been determined from the mean of the absolute deviation of
these measurements. As our spectroscopic sample also includes C-stars, this whole process was repeated with our AAOmega 
observations of six C-rich giants taken from the study of \cite{kunkel97}, details of which are reported in Table~\ref{CRVs}.

\subsection{Radial velocity measurements.}
\label{rvprec}

For the vast majority of stars, the radial velocities obtained with {\tt FXCOR} were found to be in excellent agreement with 
the values output by the $\chi^{2}$ model fitting procedure, above. The small number of exceptions can be attributed to spectra 
that are of very low signal-to-noise, datasets that are severely affected by fibre fringing or spectra of C-rich stars, which 
we discuss later on. Nonetheless, to thoroughly assess the internal precision and external accuracy of these measurements, several
 further checks have been performed. 

Firstly, a small fraction of the spectroscopically observed sample (n$\approx$175) lying in overlap regions between our 2dF/AAOmega 
field pointings were observed twice during the course of the four night run. The different velocity estimates for these objects 
have been compared to each other. Excepting the handful of stars where the discrepancy appears to be much larger than typical (ie.
of the order $\sim$10km\,s$^{-1}$), a very close correspondance is observed between the two sets of measurements, with the magnitude 
of the velocity difference for 68\% of objects being $\Delta$$v_{los}$$\le$1.9km\,s$^{-1}$. This scatter is comparable in size to the
uncertainty estimated above. 

Secondly, 17 RGB stars in the calibration cluster Melotte\,66, have been observed several times previously with AAOmega 
by other independent teams of investigators, using different fibre configurations. These sets of measurements have been 
compared to each other and after excluding a probable radial velocity variable star, discussed further below, any systematic
offsets between the various pairings of measurements were determined to be very small, $\Delta$$v_{r{_sys}}$$<$1kms$^{-1}$.
Additionally, the scatters in the velocity differences between our observations and those acquired on 08 December 2009, 24 
December 2009 and 22 April 2011, are only 1.8, 1.7 and 1.9kms$^{-1}$, respectively, consistent with very low fibre-to-fibre 
velocity differences.

\begin{table*}
\begin{minipage}{180mm}
\begin{center}
\caption{Details of the 4172 red-giants identified in our spectroscopic follow-up of sources towards the SMC. The full table is available 
in the electronic version of the article.}
\label{rgstars}
\begin{tabular}{|l|l|l|r|r|r|r|c|r|r|c|}
\hline
  \multicolumn{1}{|c|}{RA} &
  \multicolumn{1}{c|}{Dec} &
  \multicolumn{1}{c|}{2MASS\,J} &
  \multicolumn{1}{c|}{$J$} &
  \multicolumn{1}{c|}{$\delta J$} &
  \multicolumn{1}{c|}{$K_{s}$} &
  \multicolumn{1}{c|}{$\delta K_{s}$} &
  \multicolumn{1}{c|}{Helio.} &
  \multicolumn{1}{r|}{$v_{\rm model}$} &
  \multicolumn{1}{c|}{$v_{\rm helio}$} &
  \multicolumn{1}{c|}{$\delta V$} \\

  \multicolumn{1}{|c|}{hh:mm:ss.ss} &
  \multicolumn{1}{|c|}{ ${^\circ}$:$'$:$''$} &
  \multicolumn{1}{c|}{} &
  \multicolumn{1}{c|}{/mag.} &
  \multicolumn{1}{c|}{/mag.} &
  \multicolumn{1}{c|}{/mag.} &
  \multicolumn{1}{c|}{/mag.} &
  \multicolumn{1}{c|}{corr. /kms$^{-1}$} &
  \multicolumn{1}{c|}{/kms$^{-1}$} &
  \multicolumn{1}{c|}{/kms$^{-1}$} &
  \multicolumn{1}{c|}{/kms$^{-1}$} \\

\hline
  00:00:28.28 & -75:33:04.4 & 00002828-7533044 &  14.68 &   0.04 &  13.72 &   0.05 &   13.6 &  189.6 &  188.7 &    2.4\\
  00:01:43.01 & -75:35:11.9 & 00014300-7535119 &  13.94 &   0.02 &  12.94 &   0.03 &   13.6 &  156.9 &  156.2 &    2.4\\
  00:02:26.40 & -75:01:30.2 & 00022640-7501302 &  13.23 &   0.03 &  12.30 &   0.03 &   13.7 &  147.0 &  146.9 &    2.3\\
  00:02:31.86 & -75:12:37.1 & 00023186-7512371 &  14.26 &   0.03 &  13.43 &   0.03 &   13.6 &  152.3 &  153.0 &    2.1\\
  00:02:57.95 & -75:38:52.9 & 00025794-7538529 &  13.31 &   0.03 &  12.56 &   0.02 &   13.5 &   72.9 &   72.2 &    2.3\\

\hline\end{tabular}
\end{center}
\end{minipage}
\end{table*}

\begin{figure}
\includegraphics[angle=0, width=\linewidth]{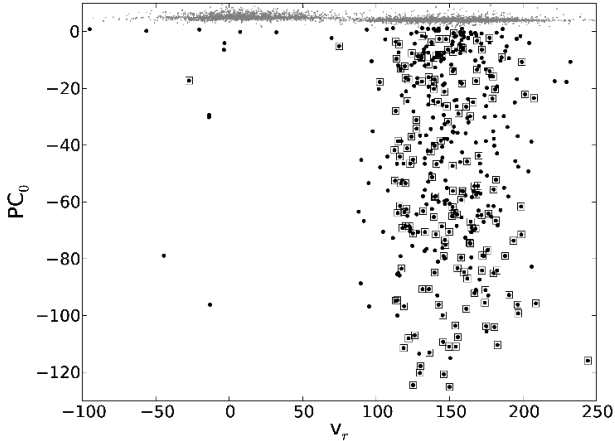}
\caption{The locations of all stars in our spectroscopic sample in the 2-d space defined by co-ordinates representing 
the strength of the C 5165\AA\ Swan band and radial velocity (grey points). Objects in our sample which have been previously 
identified as 
SMC C-rich stars are highlighted (squares). We have flagged all objects more than 5$\sigma$ below the locus defined by the 
bulk of the sample as candidate C-stars (black points).}
\label{Cstar}
\end{figure}

\begin{figure}
\includegraphics[angle=0, width=\linewidth]{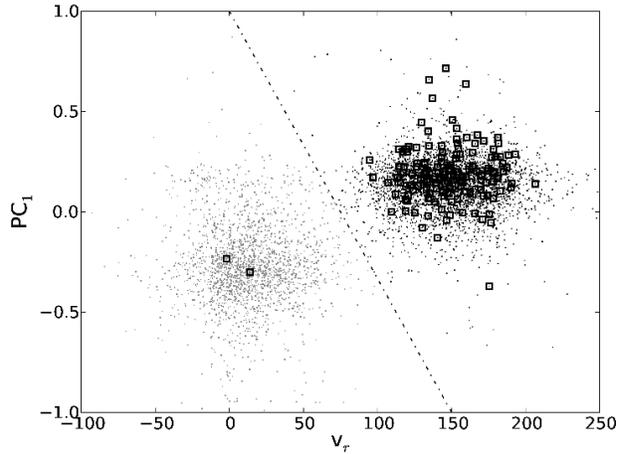}
\caption{The location of all remaining stars in our spectroscopic sample in the 2-d space defined by co-ordinates representing 
the strength of the $\lambda$5167, 5172 and 5183\AA\ Mg b lines and radial velocity (grey points). Objects in our sample which 
have been previously identified as SMC RGB stars are highlighted (squares).  The clump of stars at v$_{r}$$\approx$-15kms$^{-1}$,
PC$_{1}$$\approx$0.2 corresponds to RGB stars in NGC\,104 which lie in field 3D09FA. We have selected all objects lying to the 
right of the dot-dashed line  and with radial velocities in the range 50$\le$v$_{r}$$\le$250kms$^{-1}$ as probable red-giant members of the SMC.}
\label{Mgb}
\end{figure}

Thirdly, our AAOmega radial velocities for several Melotte 66 stars have been compared to recent measurements made with the the 
European Southern Observatory's Very Large Telescope (VLT) and Ultraviolet-visible echelle spectrograph (UVES), that are reported 
to be repeatable at the 0.5kms$^{-1}$ level \citep{sestito08}. While there are five objects in common with the AAOmega sample, star 
1346 is one of our radial velocity templates (4266) and star 1614 is flagged as a fast rotator by \cite{sestito08}. Together, the 
measurements of star 1614 suggest it is also a radial velocity variable (+17.3$\pm$2.3, +17.8$\pm$2.3, +25.6$\pm$2.3 and +57.8$
\pm$2.3kms$^{-1}$ on 08 December 2009 and 24 December 2009, 22 April 2011 and 21 October 2011, respectively). For the three remaining
stars the differences between the UVES and the AAOmega velocities (ie. v$_{\rm AAOmega}$ - v$_{\rm UVES}$) are only -0.42$\pm$2.32kms$^{-1}$
for 1493, -0.30$\pm$2.22 kms$^{-1}$ for 1785 and -0.72$\pm$2.36 kms$^{-1}$ for 2218. 

Lastly, we compared the velocity determinations for 151 SMC RGB stars common to our sample and that of \cite{harris06}. The 
latter measurements are based on observations obtained with the Magellan telescope and the multi-slit Inamori Magellan Areal Camera 
(IMACS). The scatter between the sets of measurements is determined to be approximately 13.5kms$^{-1}$, which is similar in magnitude to 
the typical uncertainties quoted by \cite{harris06}. However, a small systematic offset of +4.5kms$^{-1}$ (v$_{\rm AAOmega}$ - v$
_{\rm IMACS}$) is apparent between the results of these two studies. At face value, this seems significant but in practice it is 
probably not. The IMACS velocity measurements suffered from systematic errors of the order 10kms$^{-1}$, although substantial efforts were 
made to mitigate these \citep{harris06}.

Taking stock of the results from these comparisons, we conclude that we have met our initial goal of obtaining radial velocity 
measurements that are repeatable to better than 5kms$^{-1}$ for the vast majority of the red-giants in our sample.

\subsection{C-rich stars and field dwarfs.}
\label{pca}

An additional goal of this work is to investigate the metallicity of the intermediate age population of the SMC so the primary focus
of our study is the RGB star population. However, the spectroscopically observed sample includes various other stellar types too, such 
as field dwarfs and C-rich giants. The data obtained with the blue arm of the AAOmega spectrograph has been used to resolve these populations 
from each other. The 1500V 
data for the objects was first shifted into the rest frame using our estimates of the stellar radial velocities obtained 
from the red-arm spectrum during the model fitting process. A 50\AA\ wide section of blue arm data centered on the 
$\lambda$5167, 5172 and 5183\AA\ Mg b lines was then cut-out and normalised. These features are known to be sensitive to
surface gravity (ie. are weaker at lower surface gravities), so their observed shapes can be exploited to separate the giants 
from the field dwarfs. Additionally, the energy distributions of the C-rich stars exhibit a distinctive Swan band feature
at 5165\AA, so this wavelength range is useful for the discrimination of these objects too.

\begin{table*}
\begin{minipage}{180mm}
\begin{center}
\caption{Details of the 352 carbon rich giants that were included in our spectroscopic survey of the SMC. The full table is available in the electronic version of the article.}
\label{cstars}
\begin{tabular}{|l|l|l|r|r|r|r|c|r|r|c|}
\hline
  \multicolumn{1}{|c|}{RA} &
  \multicolumn{1}{c|}{Dec} &
  \multicolumn{1}{c|}{2MASS\,J} &
  \multicolumn{1}{c|}{$J$} &
  \multicolumn{1}{c|}{$\delta J$} &
  \multicolumn{1}{c|}{$K_{s}$} &
  \multicolumn{1}{c|}{$\delta K_{s}$} &
  \multicolumn{1}{c|}{Helio.} &
  \multicolumn{1}{r|}{$v_{\rm model}$} &
  \multicolumn{1}{c|}{$v_{\rm helio}$} &
  \multicolumn{1}{c|}{$\delta V$} \\

  \multicolumn{1}{|c|}{hh:mm:ss.ss} &
  \multicolumn{1}{|c|}{ ${^\circ}$:$'$:$''$} &
  \multicolumn{1}{c|}{} &
  \multicolumn{1}{c|}{/mag.} &
  \multicolumn{1}{c|}{/mag.} &
  \multicolumn{1}{c|}{/mag.} &
  \multicolumn{1}{c|}{/mag.} &
  \multicolumn{1}{c|}{corr. /kms$^{-1}$} &
  \multicolumn{1}{c|}{/kms$^{-1}$} &
  \multicolumn{1}{c|}{/kms$^{-1}$} &
  \multicolumn{1}{c|}{/kms$^{-1}$} \\
\hline
  00:04:57.49 & -76:25:07.7 & 00045748-7625076 &  13.33 &   0.02 &  12.19 &   0.02 &   13.4 &  162.3 &  156.9 &    6.8\\
  00:06:12.83 & -75:16:21.1 & 00061283-7516211 &  13.71 &   0.02 &  12.36 &   0.02 &   13.5 &  154.3 &  149.7 &    5.9\\
  00:08:09.42 & -75:19:06.1 & 00080942-7519060 &  13.56 &   0.02 &  12.32 &   0.02 &   13.5 &  145.5 &  139.9 &    5.5\\
  00:11:31.72 & -73:59:54.0 & 00113171-7359539 &  13.99 &   0.03 &  13.07 &   0.03 &   13.6 &  149.5 &  144.8 &    8.9\\
  00:14:58.50 & -75:07:29.4 & 00145849-7507294 &  12.90 &   0.02 &  11.39 &   0.02 &   13.3 &  167.9 &  162.5 &    5.9\\

\hline\end{tabular}
\end{center}
\end{minipage}
\end{table*}

A set of orthogonal basis vectors was constructed to represent all the blue-arm dataset sub-sections. The principal 
eigenvector formed in this process (which accounts for approximately 40\% of the variance) can be attributed to the spectral shape 
induced by the strong molecular-C absorption in the atmospheres of some stars. Comparing the locations of all the spectroscopic targets 
and the 185 objects previously identified as C-star members of the SMC and re-observed here \citep{morgan95}, 
in the 2d-space defined by this new co-ordinate and radial velocity, the C-rich stars are observed to lie well below the locus
of points de-lineated by the bulk of the sample (see figure~\ref{Cstar}). A by-eye inspection of the red-arm data
sets for a random selection of these objects, not previously catalogued as C-rich, affirms the presence of strong molecular
C absorption. A 5$\sigma$ clip has been applied to the main locus of points shown in figure~\ref{Cstar} and the 449 objects 
below have been flagged as probable C-stars.

Subsequently, the C-stars were removed from the sample and the set of basis vectors for the remaining blue-arm data was 
re-constructed. The principal eigenvector (corresponding to approximately 10\% of the variance) can now be ascribed to the 
$\lambda$5167, 5172 and 5183\AA\ Mg b lines and the variation of their form with surface gravity. The locations of all remaining 
spectroscopic targets in the 2d-space defined by this new co-ordinate and radial velocity has been compared to those of objects 
previously identified as red-giant members of the SMC and re-observed here \citep{harris06}. A cursory glance 
at figure~\ref{Mgb} reveals that the lower gravity giants are rather well separated from the field dwarfs. For example, despite having 
a relatively small separation in radial velocity from the field population, the RGB members of NGC\,104, which encroached
on one of our field configurations (3D09FA), are visible in this plot as a clump of objects at v$_{r}$$\approx$-15kms$^{-1}$, 
PC$_{1}$$\approx$0.2. The 4172 datapoints lying to the right of the line defined by PC$_{1}$= -0.013 v$_{r}$ + 1.0 (dot-dashed line in 
figure~\ref{Mgb}) and with radial velocities in the range 50$\le$v$_{r}$$\le$250kms$^{-1}$ have been associated with probable red 
giant members of the SMC. Details of these objects are listed in Table~\ref{rgstars}.

\begin{figure}
\label{rvhisto}
\includegraphics[angle=0, width=\linewidth]{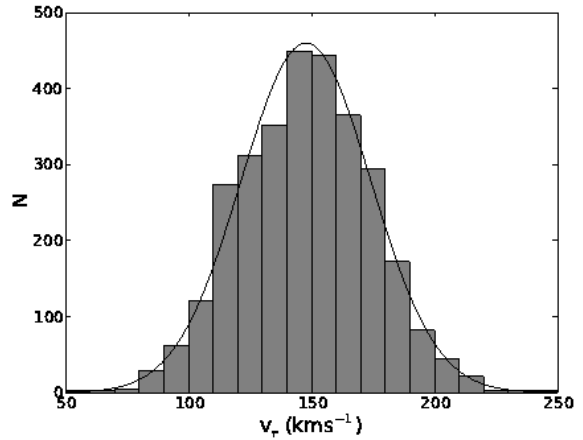}
\caption{The histogram of the radial velocity estimates for our entire sample of SMC RGB stars (grey).} 
\label{rvhisto}
\end{figure}

\begin{figure}
\label{crvhisto}
\includegraphics[angle=0, width=\linewidth]{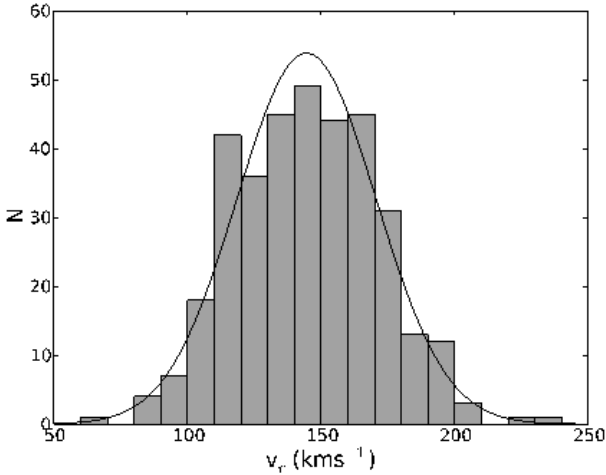}
\caption{Histogram of the radial velocities of all C-rich stars with $J$$<$14.0 identified in our study. The best fitting Gaussian function 
to the distribution is overplotted (solid black line).} 
\label{crvhisto}
\end{figure}

\begin{table*}
\begin{minipage}{90mm}
\begin{center}
\caption{A table summarising the results from fitting Gaussians to the radial velocity histograms for our spatial 
sub-samples of RGB stars.}
\label{Metals}
\begin{tabular}{lrcrr}
\hline
Field & N$_{\rm tot}$ & $v_{\rm r}$ & $\sigma_{v_{\rm r}}$ & $\Delta v_{\rm r}$\\ 
    & & /kms$^{-1}$ & /kms$^{-1}$ & /kms$^{-1}$\\
\hline

C1     & 321  & +135.6$\pm$1.3& 22.8$\pm$0.9  & -5.4$\pm$1.3     \\
C2     & 390  & +143.2$\pm$1.2& 23.2$\pm$0.8 & -0.5$\pm$1.1     \\
C3     & 586  & +148.6$\pm$1.0 & 25.1$\pm$0.8 & -0.6$\pm$1.1     \\
C4     & 333  & +158.9$\pm$1.4 & 24.6$\pm$1.0 &  +6.8$\pm$1.3   \\
3D05   & 57   & +142.2$\pm$3.4& 25.5$\pm$2.4 & +16.0$\pm$3.4     \\
3D06   & 99   & +136.0$\pm$2.7& 26.8$\pm$1.9 & +6.0$\pm$2.8      \\
3D07   & 112  & +134.9$\pm$2.3& 24.0$\pm$1.7 & -0.7$\pm$2.2     \\
3D08   & 115  & +135.5$\pm$2.1& 22.4$\pm$1.5  & -6.0$\pm$2.1     \\
3D09   & 145  & +133.8$\pm$2.2& 25.9$\pm$1.6  & -1.9$\pm$2.2     \\
3D10   & 120  & +147.9$\pm$2.3& 24.7$\pm$1.7  & -2.4$\pm$2.2     \\
3D11   & 121  & +145.4$\pm$2.2& 24.5$\pm$1.6  & -0.1$\pm$2.2     \\
3D12   & 124  & +162.0$\pm$2.3& 25.3$\pm$1.7  & +3.4$\pm$2.4     \\
3D13   & 108  & +149.5$\pm$2.4& 25.0$\pm$1.7  & -3.1$\pm$2.4     \\
3D14   & 142  & +157.3$\pm$1.8& 21.7$\pm$1.3  & -0.0$\pm$1.8     \\
3D15   & 170  & +163.8$\pm$1.9& 24.9$\pm$1.4  & +2.1$\pm$1.9     \\
3D16   & 92   & +162.6$\pm$3.1& 29.2$\pm$2.2  & -4.8$\pm$3.1     \\
5D19   & 29   & +136.3$\pm$4.4 & 23.4$\pm$3.2 & -2.3$\pm$4.5     \\
5D22   & 29   & +156.1$\pm$5.7& 30.3$\pm$4.1  &  -0.4$\pm$5.6    \\

\hline
\label{Metals}
\end{tabular}
\end{center}
\end{minipage}
\end{table*}

\begin{figure}
\label{brvhisto}
\includegraphics[angle=0, width=\linewidth]{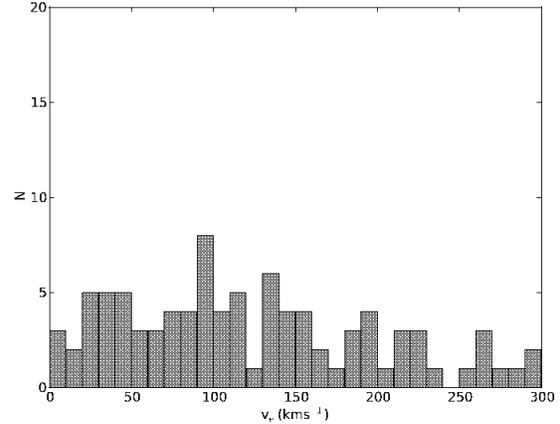}
\caption{The histogram of the radial velocities of the contaminating Galactic giant stars that meet our 
survey selection criteria, as predicted by the Besancon model.} 
\label{brvhisto}
\end{figure}

\section{Radial velocities of the SMC giants}

\subsection{The sample dominanted by RGB stars}

\cite{cioni00} determine the tip of the RGB in the SMC to lie at $J$$\approx$13.7 so by conservatively selecting the stars 
with $J$$\ge$14, which weren't flagged as C-rich in Section~\ref{pca}, we form a sample that is dominated by objects on the 
RGB and appropriate for a metallicity study. With radial velocities for 3037 unique RGB sources we are also in a strong position 
to explore the kinematics of the intermediate age stellar population across a large swathe of the Cloud. Indeed, the Besancon
model of the Galaxy \citep{robin03} reveals that only 65 contaminating Galactic giants (just 2\% of the sample) meet our SMC 
RGB star colour and radial velocity selection criteria. Since a histogram of their radial velocities (figure~\ref{brvhisto}) shows 
these to be relatively evenly spread across our parameter space, we conclude that they are unlikely to have any signficant bearing
on our subsequent analysis and conclusions. 

The typical precision of our radial velocity measurements has been discussed in Section~\ref{rvprec}. A histogram of them is 
shown in figure~\ref{rvhisto}. We have calculated both the skew ($\gamma_{1}$=0.057$\pm$0.044) and the kurtosis ($\gamma_{2}
$=-0.084$\pm$0.088) of the overall radial velocity distribution and have found both to be consistent with normality. The results
of a Kolmogorov-Smirnov test for normality (P=0.12) are also compatible with this conclusion. We have determined the mean and 
dispersion of the radial velocities to be $v_{los}$$\approx$147.8$\pm$0.5km\,s$^{-1}$ and $\sigma$$_{v_{los}}$$\approx$26.4$\pm$0.4km\,s$^{-1}$,
respectively, with the maximum likelihood estimator,

\begin{equation}
\begin{split}
\ln(p) & = -\frac{1}{2}\left( \sum_{m=1}^{N}\ln(\sigma_{m}^{2} + \sigma_{v_{los}}^{2})  + \sum_{m=1}^{N}\frac{(v_{m} - v_{r})^{2}}{(\sigma_{m}^{2} + \sigma_{v_{los}}^{2})}\right) \\
& - \frac{N}{2} \ln(2\pi)
\end{split}
\end{equation}

where $N$ is the number of RGB stars in our sample and $v_{m}$ and $\sigma_{m}$ correspond to the individual radial velocity measurements and 
their uncertainties, respectively \citep[e.g.][]{walker06p}. The derived parameters are broadly in agreement with the results of previous, 
independent, studies of the intermediate age stellar populations of the SMC \citep[e.g.][]{harris06, hatzidimitriou97,hatzidimitriou93, hardy89, dopita85}.

\subsection{Carbon stars}

We also obtained spectroscopic data for several hundred candidate C-rich SMC giants (see Section~\ref{pca}). These objects, details of which are listed 
in Table~\ref{cstars}, span the full magnitude range of our study from $J$=12.0--15.2 but we have restricted our kinematic analysis to the 352 stars with
$J$$<$14.0 that are located in the canonical red AGB wing of the SMC colour-magnitude diagram (figure~\ref{CMD}). Theoretical models suggest that the lowest 
luminosity C-rich giants may be formed in close binary systems \citep[e.g.][]{marigo99}. Additionally, the objects in this sub-sample were assigned uniformly
higher priorities for spectroscopic follow-up than the stars with $J$$>$14.0.

As discussed briefly in Section~\ref{pca}, the radial velocities of the objects flagged as C-rich were determined by cross-correlating their spectra against
those of six C-rich SMC giants previously investigated by \cite{kunkel97}. These authors noted their radial velocities to be systematically shifted to the 
blue by about 6 kms$^{-1}$ with respect to the measurements of \cite{hardy89}. We observe an offset of similar magnitude and direction between the velocities
we obtained from cross-correlation and the estimates output by our model fitting procedure. While we caution that the synthetic spectra used here were hardly
ideal for matching to C-stars, we have applied an offset of +5kms$^{-1}$ to our measurements to bring them into closer agreement with both the system of 
\cite{hardy89} and our model based estimates. Following the approach taken with the RGB stars we have examined both the skew ($\gamma_{1}$=0.109$\pm$0.131) and 
the kurtosis ($\gamma_{2}$=-0.109$\pm$0.261) of the C-star radial velocity distribution. We have found both of these parameters, together with the results of a 
Kolmogorov-Smirnov test (P=0.833), to be in accord with normality (figure~\ref{crvhisto}). We have used the the maximum likelihood estimator above to determine
the mean and dispersion of this distribution to be $v_{los}$$\approx$149.6$\pm$1.4km\,s$^{-1}$ and $\sigma_{v_{los}}$$\approx$26.1$\pm$1.0km\,s$^{-1}$, which are in 
accord with the parameters of the RGB star ensemble.


\subsection{Space motion of the SMC centre of mass.}
\label{COMM}

To search the RGB radial velocities for evidence of systematic variation with position on sky, we have initially split the measurements up into 18 sub-samples, 
each corresponding to a distinct 2dF field pointing. Basic parameters (e.g. the mean and the dispersion) for the radial velocity distributions of these fields have
been obtained by applying the above likelihood statistic, under the assumption they too, are gaussian. The results from this procedure are shown graphically in 
figure~\ref{rvhistosp} and are listed in Table~\ref{Metals}. These reveal an overall radial velocity gradient across the sample of about +7kms$^{-1}$\,deg$^{-1}$, 
in an approximately NW-SE direction. Our much smaller sample of C-stars also reflects this trend (see figure~\ref{crvhistosp} and Table~\ref{cstars}).

\begin{figure}
\label{rvhistosp}
\includegraphics[angle=0, width=\linewidth]{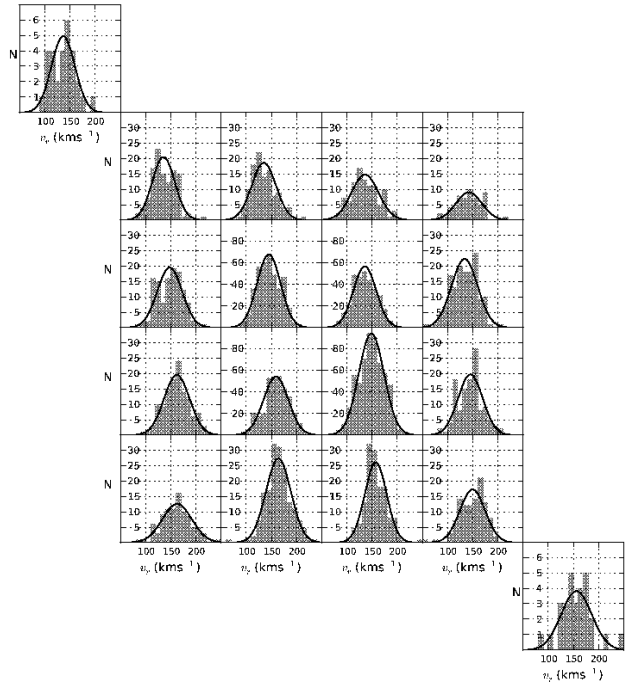}
\caption{Histograms used to explore the spatial dependence of the mean radial velocity and its dispersion in our sample of SMC RGB stars. 
The best fitting Gaussian function to each distribution is overplotted (solid black line).} 
\label{rvhistosp}
\end{figure}

\begin{figure}
\label{velofield}
\includegraphics[angle=0, width=\linewidth]{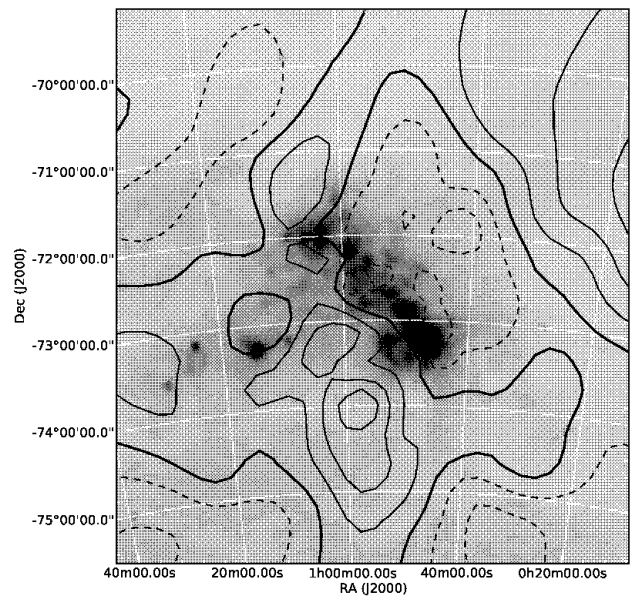}
\caption{A 6.0$^{\circ}$$\times$6.5$^{\circ}$ contour map of the RGB star galaxy rest frame velocity surface after subtraction of a solid body 
model of the centre-of-mass proper motion of the SMC (see text for details). The velocity data around each grid point were smoothed
using an adaptive gaussian kernel with a width corresponding to one third of the distance to the 200th closest star. The contours correspond to
steps of 4kms$^{-1}$ (heavy line 0.0 kms$^{-1}$, dashed lines negative velocities).} 
\label{velofield}
\end{figure}

\begin{figure}
\label{velofieldkernel}
\includegraphics[angle=0, width=\linewidth]{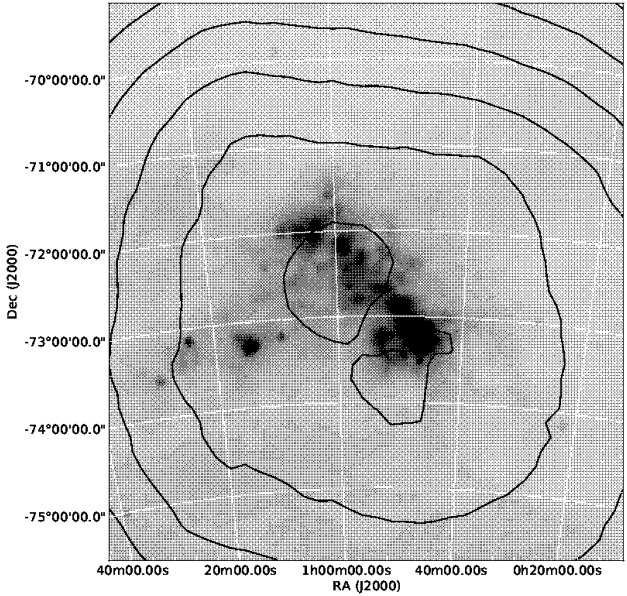}
\caption{A 6.0$^{\circ}$$\times$6.5$^{\circ}$ map of the width of the smoothing kernel used to produce figure~\ref{velofield}. The contours correspond to
steps of 10 arcmin, where the inner most contour corresponds to 10 arcmin.} 
\label{velofieldkernel}
\end{figure}

\begin{figure}
\label{vprob}
\includegraphics[angle=0, width=\linewidth]{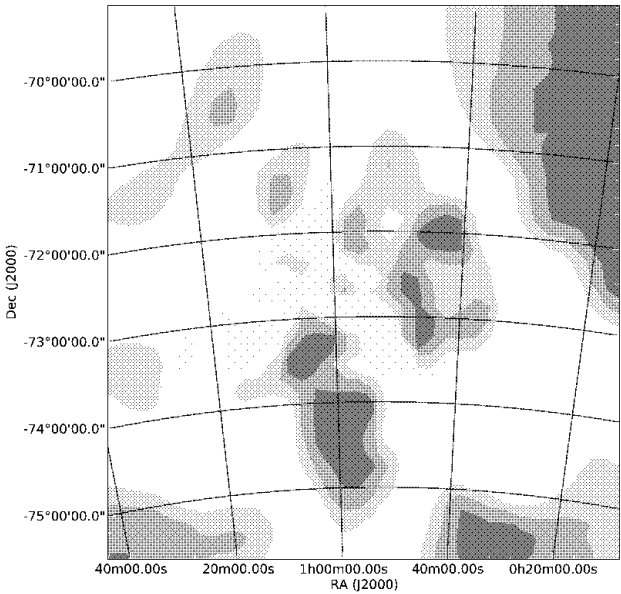}
\caption{A 6.0$^{\circ}$$\times$6.5$^{\circ}$ contour map of the statistical likelihood that the features in the galaxy rest frame velocity surface are not merely 
due to chance (P=0.95--0.99 in light gray, P=0.99--0.999 in gray and P$>$0.999 in dark gray).} 
\label{vprob}
\end{figure}

As discussed by previous investigators \citep[e.g.][]{vandermarel02}, a substantial contribution to our observed radial velocity gradient 
may stem from the expected variation in the line-of-sight velocity component of the Cloud's space velocity across our extensive 
survey area. A 
number of estimates of the proper motion of the SMC centre-of-mass have been published since the RGB star study of \cite{harris06} and 
these have substantially reduced uncertainties compared to similar earlier work. The most recent is based on three epochs of Hubble Space
Telescope imaging and has measured $\mu_{\alpha cos \delta}$=0.772$\pm$0.063mas\,yr$^{-1}$ and $\mu_{\delta}$=-1.117$\pm$0.061mas\,yr$^{-1}$ 
\citep{kallivayalil13}. We have used this information in conjunction with a kinematical model for solid body rotation \citep[e.g.][]{vandermarel02} 
to assess the impact of the Cloud's tangential motion on our measured values. We have neglected for now any contribution to the velocities from a 
putative disk structure, and, for consistency with the work of \cite{kallivayalil13}, we have assumed initially that the SMC center of mass 
is coincident with the HI kinematic center \citep{stanimirovic04}. This model was matched to the measured radial velocities of the individual
RGB stars by locating the global minimum of a $\chi^{2}$ goodness-of-fit statistic.
We allowed the model parameters $v_{sys}$, $v_{t}$ and  $\Theta_{\rm t}$ (in the notation of \citet{vandermarel02}, respectively the systemic 
velocity, the tangential velocity and the position-angle of the tangential velocity, east from north) to vary freely in this process. We assumed 
for now an intrinsic velocity dispersion of $\sigma_{v_{los}}$=25 kms$^{-1}$, which is compatible with both the typical values we measure for the 
sub-samples across the cluster and the results of earlier studies of the intermediate age stellar population of the Cloud. 

We find this basic model can provide a reasonable match to the data with a reduced-$\chi^{2}$$\approx$1 for parameter values of $v_{sys}$=147.5$\pm$0.5kms$
^{-1}$, $v_{t}$=416.8$\pm$23.0kms$^{-1}$ and $\Theta_{\rm t}$=152.1$\pm$2.9$^{\circ}$. The errors quoted here were obtained via a bootstrap with random 
replacement approach. The broad agreement between these parameters and the values inferred from the most recent estimate of the SMC centre of mass proper 
motion, $v_{t}$=386$\pm$21kms$^{-1}$ and $\Theta_{\rm t}$=145.4$\pm$2.6$^{\circ}$ (assuming a distance modulus of ($m$-$M$)$_{0}$=18.90) and our determination of
the radial velocity,  $v_{sys}$=147.7$\pm$0.5kms$^{-1}$, argues that any manifestation of systemic rotational motion in the RGB star kinematics has an amplitude 
well below the velocity dispersion of this population.

\begin{figure}
\label{crvhistosp}
\includegraphics[angle=0, width=\linewidth]{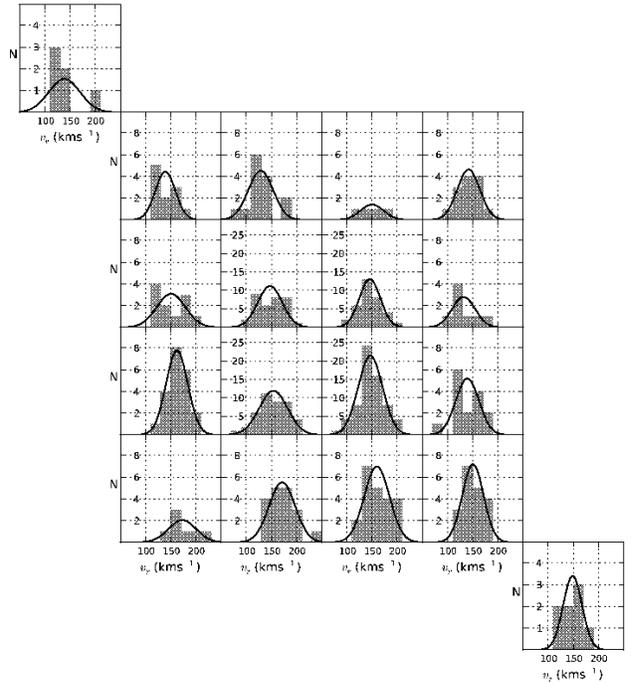}
\caption{Histograms used to explore the spatial dependence of the C-star radial velocities. The best fitting Gaussian function to each distribution
is overplotted (solid black line).} 
\label{crvhistosp}
\end{figure}

\begin{table}
\begin{minipage}{90mm}
\begin{center}
\caption{A table summarising the results from fitting Gaussians to the radial velocity histograms for our spatial 
sub-samples of C-stars.}
\label{cstars}
\begin{tabular}{lrccr}
\hline
Field & N$_{\rm tot}$ & $v_{\rm r}$ & $\sigma_{v_{\rm r}}$ & $\Delta v_{\rm r}$\\ 
    & & /kms$^{-1}$ & /kms$^{-1}$ & /kms$^{-1}$\\
\hline

C1     & 34   & +145.7$\pm$3.7 & 20.8$\pm$2.8  & +2.9$\pm$3.7 \\
C2     & 34   & +147.3$\pm$4.3 & 24.3$\pm$3.2  & +1.3$\pm$4.3 \\
C3     & 65   & +146.7$\pm$3.1 & 24.2$\pm$2.3  & -2.4$\pm$3.0 \\
C4     & 42   & +154.0$\pm$4.4 & 28.0$\pm$3.2  & +2.3$\pm$4.3 \\
3D05   & 13   & +141.8$\pm$6.6 & 22.4$\pm$4.8  & +15.4$\pm$6.9 \\
3D06   & 4    & +150.0$\pm$12.5 & 22.6$\pm$6.7 & +17.8$\pm$14.3 \\
3D07   & 14   & +128.8$\pm$6.9 & 24.7$\pm$5.1  & -7.0$\pm$6.4 \\
3D08   & 11   & +139.3$\pm$6.4 & 19.8$\pm$4.7  & -2.9$\pm$7.3 \\
3D09   &  8   & +132.0$\pm$8.6 & 22.8$\pm$6.4  & -5.6$\pm$8.4 \\
3D10   & 11   & +150.7$\pm$9.0 & 28.5$\pm$6.5  & +1.0$\pm$8.9 \\
3D11   & 15   & +139.5$\pm$6.3 & 23.1$\pm$4.6  & -5.8$\pm$6.2 \\
3D12   & 21   & +162.3$\pm$5.0 & 21.6$\pm$3.7  & +4.7$\pm$5.1 \\
3D13   & 19   & +150.6$\pm$5.1 & 21.1$\pm$3.8  & +0.0$\pm$5.1 \\
3D14   & 22   & +159.7$\pm$5.6 & 25.1$\pm$4.0  & +5.0$\pm$5.5 \\
3D15   & 18   & +171.4$\pm$6.4 & 26.0$\pm$4.7  & +10.9$\pm$6.1 \\
3D16   & 7   & +173.0$\pm$10.9 & 27.2$\pm$8.0 & +5.8$\pm$11.2 \\
5D19   & 6   & +138.7$\pm$13.5 & 31.2$\pm$7.5 & -2.2$\pm$14.6 \\
5D22   & 8   & +148.7$\pm$7.2 & 18.8$\pm$5.3  & -7.7$\pm$7.0 \\

\hline
\label{cstars}
\end{tabular}
\end{center}
\end{minipage}
\end{table}

\subsection{Main trends in the velocity field of the red-giant sample}
\label{vsub}

To reveal any more subtle velocity structures within our dataset, the predictions of our basic kinematical model have been subtracted from our measurements. 
Subsequently, we constructed a surface of the velocities in the rest frame of the SMC galaxy (GRF) for our survey area by estimating this parameter at a series of 
regularly spaced grid points in RA and declination (every 10 arcmin), using a bi-variate gaussian smoothing kernel, with an adaptive width corresponding to 
one third the distance to the 200th closest star, to weight the individual measurements \citep[e.g.][]{walker06}. A contour plot of this surface is displayed 
in figure~\ref{velofield}  and a map of the width of the smoothing kernel is shown in figure~\ref{velofieldkernel}. No gradient in the red-giant velocity field is obvious along the major axis of the SMC Bar. This is consistent with the results of
most previous kinematical studies of the intermediate and old populations of the SMC. For example, \cite{dopita85} reported a lack of organised structure in 
the velocities of 44 planetary nebulae located largely along the Bar, while both \cite{hatzidimitriou97} and \cite{hardy89} found no evidence of systemic 
rotation in their radial velocities of modest sized samples of C-stars.

Nonetheless, figure~\ref{velofield} reveals a rather striking dipole-like velocity pattern within roughly the central 10 deg$^{2}$ that has a major axis 
almost perpendicular to the SMC Bar. To the NW side of the Bar, negative GRF velocities at $v$$<$-5kms$^{-1}$ predominate, while immediately to the SE, positive 
velocities extend to $v$$>$+10kms$^{-1}$. An analysis of several thousand simulated velocity datasets,  which were generated by randomly re-assigning the GRF
velocities to the positions of our sample stars, indicates that this signal is statistically significant (figure~\ref{vprob}).  The implied velocity gradient here is similar in magnitude and direction to that induced by the transverse motion of the Cloud center-of-mass as estimated recently from proper motion measurements 
of the inner regions of the SMC. This gradient could be largely accounted for if these astrometric measurements were systematically underestimated by at least 
50\% in both RA and declination. However, while the most recent determination of the transverse motion of the Cloud is smaller than most previous estimates, the 
reduction does not amount to 50\%, so it seems unlikely that the observed effect is due to grossly inaccurate astrometry.
  
\cite{hardy89} measured a larger mean velocity (160.7$\pm$5.6kms$^{-1}$) for a sample of C-stars which they ascribed to the SMC Wing but 
their field appears to be coincident with one of the zones of positive GRF velocity indicated by the red-giant stars, adjacent to the SE edge of the bar. A 
contour plot of the velocity surface derived from our sample of several hundred C-stars (constructed following the above procedure) has a broadly similar pattern 
to the corresponding RGB star map but hints at larger velocities much further out in the Wing region towards the eastern limit of our survey (figure~\ref{Cvelofield}).

\begin{figure}
\label{Cvelofield}
\includegraphics[angle=0, width=\linewidth]{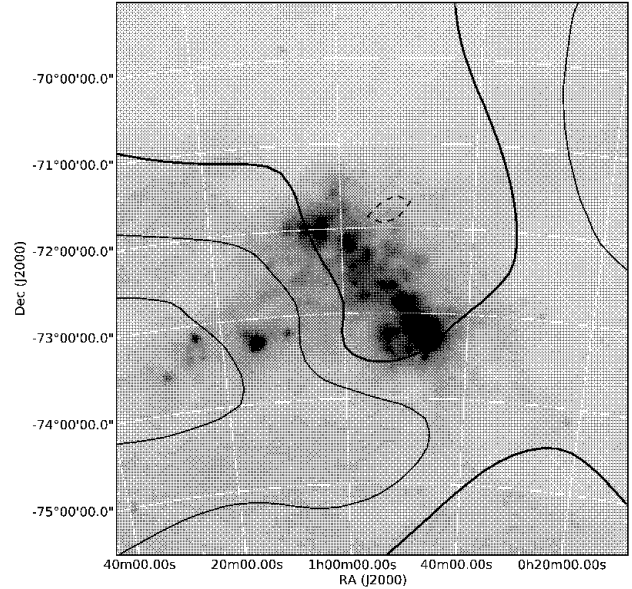}
\caption{A 6.0$^{\circ}$$\times$6.5$^{\circ}$ contour map of the carbon star GRF velocity surface after subtraction of a solid body 
model of the centre-of-mass proper motion of the SMC. Other details of this plot are the same as in figure~\ref{velofield}.} 
\label{Cvelofield}
\end{figure}

In contrast to our work, \cite{harris06} found evidence of a velocity gradient of +8.3kms$^{-1}$\,deg$^{-1}$ at PA$\approx$23$^{\circ}$, east of north, in their 
investigation of 2046 red-giants drawn from 12 fields located mainly around the SMC Bar. Given the apparent discrepancy between their result and that of the 
present work, we have explored their conclusion further. Considering the somewhat limited extent of their survey in the NW-SE direction, it is conceivable that
a comparatively shallow gradient along this axis may have been concealed by their larger measurement uncertainties. Alternatively, the position-angle may have 
been incorrectly referenced (e.g. from due E, rather than due N), with the value quoted by \cite{harris06} perhaps in error by -90$^{\circ}$. We have re-evaluated 
the direction of the steepest velocity gradient in our RGB ensemble using a sub-sample of 1038 stars from roughly the same region of sky as the \cite{harris06}
objects. We have added random velocity offsets, drawn from a gaussian distribution with a mean of 0.0 kms$^{-1}$ and a width of 10 kms$^{-1}$, to our measurements
to ensure that the uncertainties for stars in the two samples have similar magnitudes. Subsequently, we have calculated half the difference between the mean 
velocities of our red-giant stars on either side of a line bi-secting the sub-sample, through the optical center of the Cloud, stepping the position-angle of this
line in 10$^{\circ}$ increments from 0$^{\circ}$ to 360$^{\circ}$ \citep[e.g.][]{mackey13}. This entire process was repeated 100 times and, despite the smaller sample 
size, in every case the parameters of the resulting velocity curve were consistent with the steepest gradient lying along an approximately NW to SE direction 
(peak amplitude of between 4 -- 6kms$^{-1}$ and a position-angle at the maximum differential velocity in the range PA$\approx$25--45$^{\circ}$). Next, we applied the 
above procedure to their 2046 stars and found the differential velocity curve to have a peak amplitude of approximately 6kms$^{-1}$ and a position-angle at the
maximum value of PA$\approx$45$^{\circ}$. The broad agreement between the values obtained from their dataset and ours suggests that the likely explanation for the 
disparity between the two results is that the position-angle quoted in \cite{harris06} was inadvertently referenced from E.

Our contour plot of the RGB star velocity surface also reveals a sizeable kinematic structure towards the NW of our survey area, where GRF velocities reach values 
of $v$$>$+10kms$^{-1}$. This region appears linked to the positive velocities SE of the central SMC via the southern end of the Bar. In the prominent eastern, 
Wing region of the Cloud, the velocity field displays no overwhelming trend. The lack of a strong, positive signal in the NE zone
of our RGB star velocity field contrasts with the findings of kinematical studies of the HI gas \citep[e.g.][]{stanimirovic04}, which were taken to be indicative of systemic 
rotation around the minor axis of the Bar. In fact, there is a substantial pool of low velocities further to the NE, which is also evident as a secondary peak (centered 
on the 115-130kms$^{-1}$ bin) in the histograms for fields 3D7, 3D8 and 3D10 (e.g. see figure~\ref{rvhistosp}). \cite{depropris10} have also noted the velocity distribution 
of the RGB stars to the east (and the south) of the SMC center to be bi-modal, but their reported peaks are at approximately 160kms$^{-1}$ and 200kms$^{-1}$, somewhat larger 
than the values we observe. This region of our survey encompasses a small sample of red clump stars spectroscopically examined by \cite{hatzidimitriou97} and with which 
they identified a positive correlation between velocity and distance. \cite{nidever13} have recently identified two relatively distinctive intermediate-age stellar structures, 
in terms of distance (approximately 55 kpc and 67kpc), projected several degrees to the east and north of the Cloud.

\subsection{Old/intermediate versus young stellar population}
\label{OVY}

To gain further insight, we have compared our results to those from another large-scale 2dF based kinematical study of the stellar population of the Cloud. 
\cite{evans08} observed 2045 massive stars and also identified a trend of increasing radial velocity from NW to SE, across the bar. It was concluded 
that this velocity gradient of approximately +25kms$^{-1}$deg$^{-1}$ at a position-angle PA$\approx$126$^{\circ}$ could not be attributed solely to variation in 
the viewing angle of the SMC's centre-of-mass motion. In accord with earlier investigations \citep[e.g][]{maurice87}, they found the stars in the Wing 
to have significantly larger velocities than those of the Bar ($v_{los}$$\approx$195kms$^{-1}$). Interestingly, most of the objects in this region turned out to be 
amongst the earliest spectral-types surveyed in their work (ie. O and early-B), which concurs with the finding of \cite{cignoni13} that there has been a 
substantial increase in the rate of star formation here within the last 200Myr. However, the majority of the later-type 
supergiants that were observed concentrate in two main elongated aggregates to the west of $\alpha$=01$^{\rm h}$12$^{\rm m}$, one extending NE-SW along the Bar (W)
and the other, from the NE end of the Bar, south along $\alpha$=01$^{\rm h}$05$^{\rm m}$ (E) (figure~\ref{evans}). The 2dF spectroscopic fibre allocation process
could conceivably have led to some apparent differences in the spatial distributions of these massive star populations but there is some evidence of the E aggregate 
in a density isopleth contour plot for SMC stars with ages in the range 0.1Gyr $<$ $\tau$ $<$ 0.3Gyr and 0.3Gyr$<$ $\tau$ $<$ 1Gyr. This structure is not apparent 
in the corresponding plot for the youngest objects with $\tau$ $<$ 0.1Gyr \citep[figure 4 of ][]{belcheva11}. There are also hints of this bi-modality in the spatial 
distribution of SMC star clusters with ages less than 3.5 Gyr \citep[e.g. figure 15][]{rafelski05}.

\begin{figure}
\label{evans}
\includegraphics[angle=0, width=\linewidth]{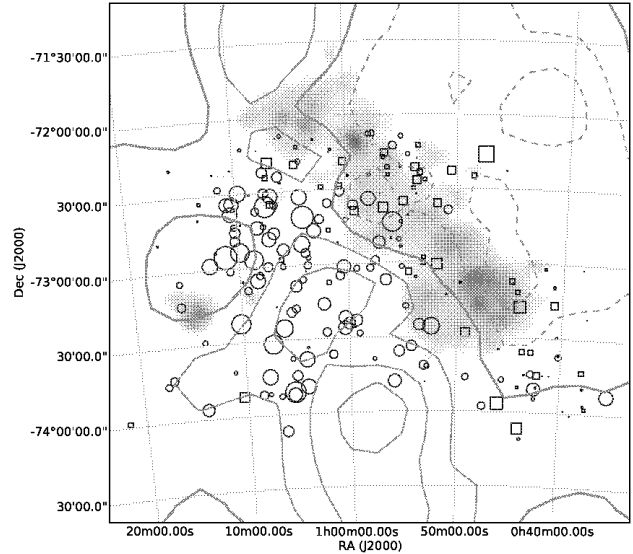}
\caption{A 3.5$^{\circ}$$\times$3.5$^{\circ}$ zoom in on figure~\ref{velofield}. The entire F/G supergiant sample from {\protect \cite{evans08}} is overplotted
with those stars having positive GRF velocities represented as circles and those with negative velocities shown as squares. The size of the 
symbols scale linearly with the magnitude of the residuals, the NW and W most stars having -83 kms$^{-1}$ and +55 kms$^{-1}$, respectively. The division of negative 
and positive F/G supergiant velocities follows closely the zero contour delineated by the RGB stars.} 
\label{evans}
\end{figure}

\begin{figure}
\label{evans_rotc}
\includegraphics[angle=0, width=\linewidth]{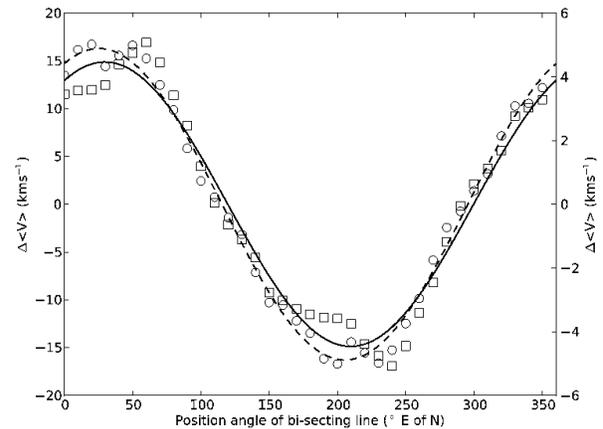}
\caption{The differences in the mean radial velocities of the F/G supergiant (circles; scaling on left-hand axis) and RGB (squares; scaling on right-hand axis) star samples on either side of a bi-secting line passing 
through $\alpha$=00$^{h}$ 53$^{m}$, $\delta$=-72$^{\circ}$ 50$^{\prime}$, as a function of position-angle. The best fitting sinusoids to these data are overplotted
(F/G supergiants, dashed line, RGB stars, solid line).} 
\label{evans_rotc}
\end{figure}

\cite{evans08} advised that their 2dF based velocity measurements were likely to be marginally over-estimated (by a mean of +10kms$^{-1}$). The results of 
our examination (ie. cross-matching with the catalogues of \citealt{maurice87, mathewson88, ardeberg79}) point to a systematic error that is somewhat dependent on spectral-type. 
While it maybe as large as 
+20kms$^{-1}$ for the O and early-B objects, it appears to be smaller than +10kms$^{-1}$ for the latest stars in the study. Consequently, in performing a direct comparison 
between the radial velocities of our red-giants and the massive stars, we have worked with only their F/G supergiants. After considering, as above, the relative space 
motion of the Cloud's centre of mass, we determine that these two aggregates of massive stars, E and W as discussed above, are dominated by positive and negative GRF velocities,  
respectively. These groupings appear to loosely correspond to the location and sign of the main velocity zones we observe in the RGB star velocity map (figure~\ref{evans}).
Figure~\ref{evans_rotc} shows the differences in the mean radial velocities of both the F/G supergiants and a sub-sample of our red-giant stars, drawn from roughly 
the same region of sky, on either side of a bi-secting line that passes through the optical center of the Cloud, \citep[$\alpha$=00$^{h}$ 53$^{m}$, $\delta$=-72$^{\circ}$ 50$^{\prime}$][]{devaucouleurs72}, as a function of this line's position-angle. 
The sinusoidal-like forms of these curves have distinct amplitudes ($\Delta$$<$${v}_{r}$$>$$^{max}$ $\approx$ 16 kms$^{-1}$ and $\Delta$$<$${v}_{r}$$>$$^{max}$ $\approx$ 4.5 kms$^{-1}$) but 
have very similar phases (PA$\approx$26$^{\circ}$ and PA$\approx$30$^{\circ}$). This conclusion is not changed significantly if, like \cite{evans08}, we had adopted 
$\alpha$=01$^{h}$ 00$^{m}$, $\delta$=-73$^{\circ}$ 00$^{\prime}$ as the center. The inferred velocity gradients perpendicular to these position-angles are +20.0$\pm$0.8kms$^{-1}$ 
deg$^{-1}$ and +6.1$\pm$0.1kms$^{-1}$ deg$^{-1}$ for the F/G supergiants and the red-giants stars, respectively. The former can easily account for the small slope of +6.5 kms$^{-1}$ 
deg$^{-1}$ measured in the F/G star GRF velocities along a PA$\approx$60$^{\circ}$, the direction of the steepest velocity gradient observed in the HI gas \citep{stanimirovic04}. 
It also appears plausible that the slope of +10kms$^{-1}$deg$^{-1}$ along this direction reported by \cite{evans08} is merely a manifestation of the gradient they identified along 
a PA$\approx$126$^{\circ}$.

Despite some marked differences between the kinematics of the intermediate/old and the massive stellar populations in the SMC (e.g. the lack of an obvious 
Wing related structure in the complete red-giant sample), there appear to be several similarities (e.g. both display a velocity gradient along a NW-SE direction).
We emphasise here that HI maps of the Cloud also provide some weak evidence of a velocity gradient extending from NW to SE, at least across the southern portion of
the Bar \citep[see figure 5 of ][]{stanimirovic04}.

\section{Interpretation and discussion}
\label{s6}

\subsection{Systemic rotational motion $?$}

\begin{figure}
\label{velofieldmp}
\includegraphics[angle=0, width=\linewidth]{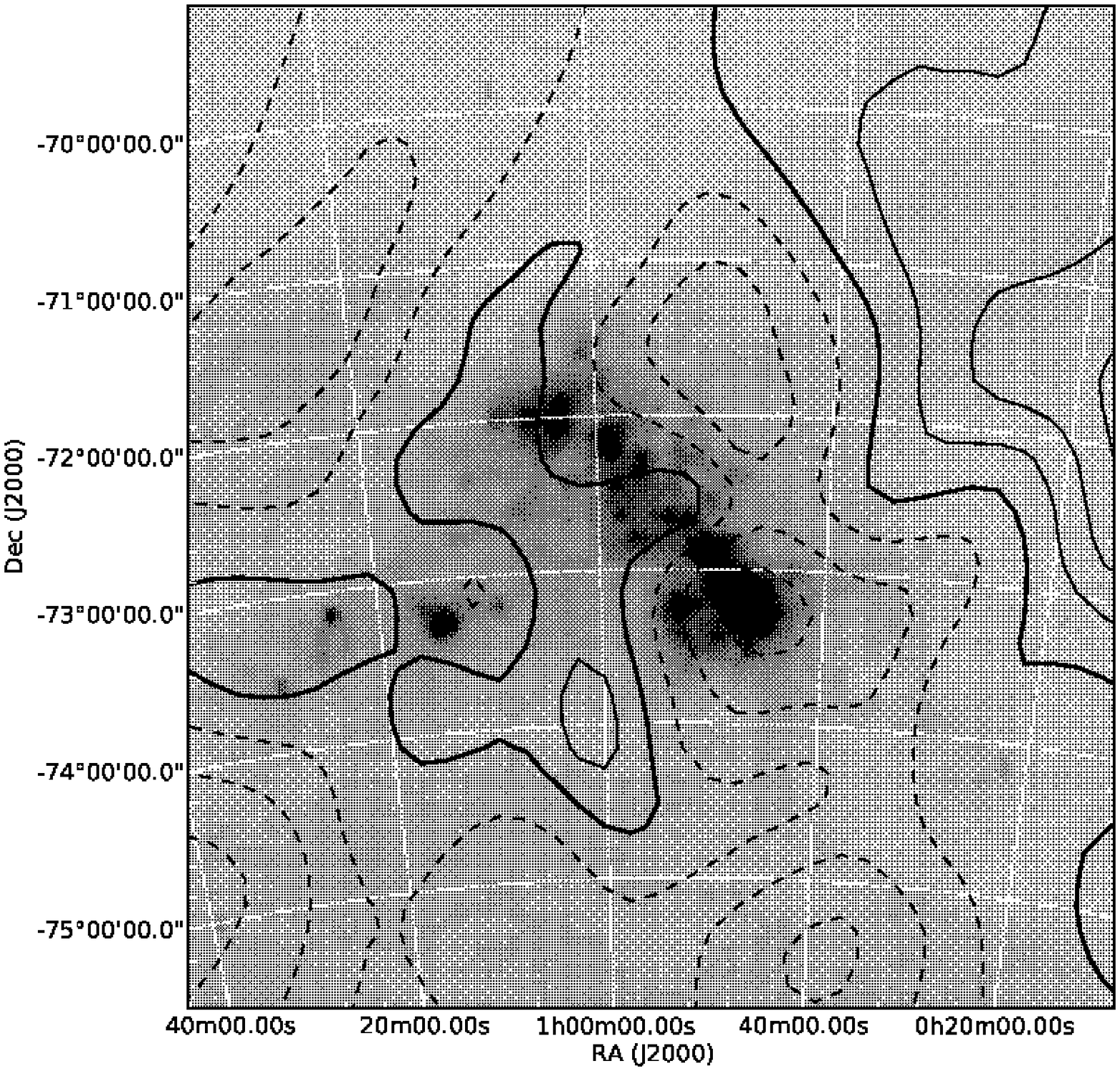}
\caption{A 6.0$^{\circ}$$\times$6.5$^{\circ}$ contour map of the GRF velocity surface for the lowest metallicity quartile of our red-giant star sample after
subtraction of a solid body model of the cente-of-mass proper motion of the SMC. Other details of this plot are the same as in figure~\ref{velofield}.} 
\label{velofieldmp}
\end{figure}

\begin{figure}
\label{velofieldmr}
\includegraphics[angle=0, width=\linewidth]{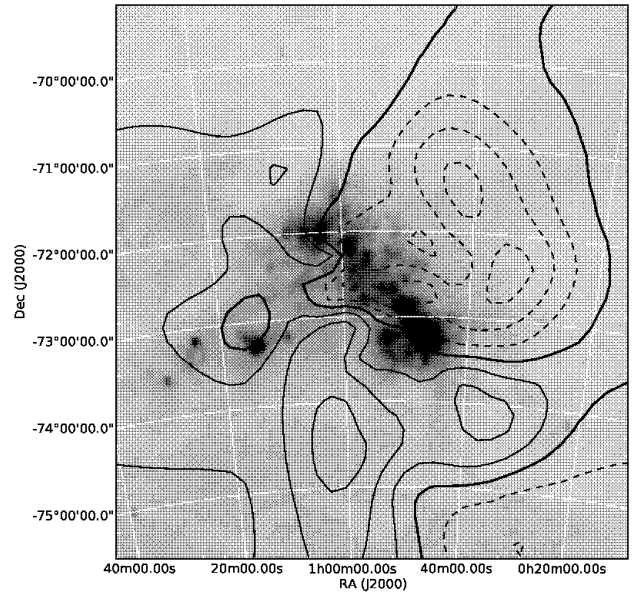}
\caption{A 6.0$^{\circ}$$\times$6.5$^{\circ}$ contour map of the GRF velocity surface for the highest metallicity quartile of our red-giant star sample
after subtraction of a solid body model of the center-of-mass motion of the SMC. Other details of this plot are the same as in figure~\ref{velofield}. } 
\label{velofieldmr}
\end{figure}

We have split our red-giant sample into quartiles on the basis of their metallicities. Considering the form of the SMC age-metallicity relation and its relative 
invariance across the galaxy \citep[e.g.][]{noel09,piatti12,cignoni13}, this is effectively a sub-division of the stellar population in age, with the more metal 
rich stars typically being significantly younger (e.g. Dobbie et al. 2014). The metallicity of each red-giant was determined from measurements of the equivalent 
widths of the CaII triplet lines and full details of this process are provided elsewhere (Dobbie et al. 2014). Subsequently, we have undertaken a comparison of the
kinematics of the upper and the lower quartiles ([Fe/H]$>$-0.86 and [Fe/H]$<$-1.15, respectively)
to gain greater understanding of the structure of the SMC. For each subsample we have constructed a contour plot of the radial velocities remaining after accounting 
for the center-of-mass motion of the Cloud (figures~\ref{velofieldmp} and \ref{velofieldmr}). We have also determined the line-of-sight velocity and dispersion profiles 
of these sub-samples and the F/G supergiants, along a line through the optical center at PA=120$^{\circ}$, which we found in Section~\ref{OVY} aligned with the maximum velocity gradient (figure~\ref{rcurve}).
The latter population shows substantial changes in velocity along this axis, reaching approximately +30kms$^{-1}$ at the south-eastern limit of the \citeauthor{evans08} 
survey coverage (corresponding to a projected distance from the SMC center of approximately 1kpc). Away from the north-western limit of our survey region (discussed further
below), our metal poor and presumed older red-giants display a line-of-sight velocity profile that is effectively flat ($\chi^{2}$=7.9 for 9 degrees of freedom). 
The line-of-sight velocity profile of our metal rich and generally younger red-giant sub-sample displays a significant gradient, albeit less pronounced than that of the F/G
supergiants, reaching 8-10kms$^{-1}$, 1-1.5 kpc from the centre. Considering that it is a combination of random and systemic rotational motions of stars around a galaxy which
act to balance the gravitational potential of the system, it is interesting that these latter stars also have a significantly smaller mean line-of-sight velocity dispersion,
$\sigma_{v_{los}}$$\approx$22.3$\pm$0.6kms$^{-1}$, than the older metal poor red-giants, $\sigma_{v_{los}}$$\approx$26.1$\pm$0.7kms$^{-1}$ (at angular distances$>$-1.5$^{\circ}$). As stars age, repeated gravitational encounters increase the random component of their mean velocities \citep[e.g.][]{binney08}. The increased velocity dispersion and lower apparent rotation velocity of the metal-poor RGB stars relative to the metal-rich portion of the sample supports the contention that an age-metallicity relation exists in the SMC.

The inferred gradient of 30-35kms$^{-1}$deg.$^{-1}$ in the F/G star population is comparable in magnitude to that predicted by the \cite{bekki09} simulations of the 
SMC/LMC/Galaxy interaction, although it is significantly less than anticipated by the N-body calculations of \cite{gardiner96}. The intermediate and old stellar populations 
are also anticipated to display a velocity gradient but the models attribute this largely to streaming motions along the tidal structures of the Cloud \citep[e.g.][]{diaz12}. 
However, it is not immediately obvious why our two sub-samples display different line-of-sight velocity profiles since strong tidal forces from an interaction with the LMC (or
the Galaxy) within the last few 100Myr would have presumably affected all our intermediate-age red-giant stars. 

In view of the coincidence of the position-angles of the velocity gradients in our red-giant and the F/G supergiant samples and our observation that the generally younger 
red-giants with greater metallicities exhibit a larger systemic motion and a lower mean velocity dispersion, we conjecture that the velocity pattern we detect is related to 
systemic rotation of disk-like structure. This notion is supported by the intriguing alignment of the Cloud's coherent magnetic field almost perpendicular to the Bar, as 
mapped through the polarisation of radio continuum emission and optical starlight \citep[e.g.][]{mathewson70, mao08}. The magnetic fields of both the Galaxy and the LMC are 
observed to trace the spiral structure of their disks. \cite{wayte90} previously concluded from his analysis of the SMC magnetic field that the Bar is not a true bar structure 
and instead, interaction with the LMC prompted an extension of the Cloud towards the NE and the formation of the Wing.

\begin{figure}
\label{rcurve}
\includegraphics[angle=0, width=\linewidth]{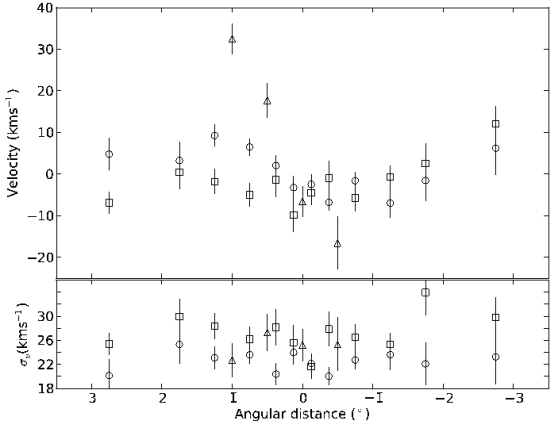}
\caption{Velocity (upper) and velocity dispersion (lower) profiles along PA=120$^{\circ}$ for the metal poor red-giants (squares), metal rich red-giants (circles) and 
the F/G supergiants (triangles).}
\label{rcurve}
\end{figure}

\subsection{Modelling the velocities as a putative SMC disk}

To investigate the above possibility, we have attempted to replicate the measured radial velocities of the metal-rich red-giant stars by adding a disk component 
to the kinematical model of Section~\ref{COMM}, following the formalism of \cite{vandermarel02}. This time we have assumed a distance to the SMC of 60kpc and 
adopted a tangential velocity of $v_{t}$=386kms$^{-1}$ at position-angle $\Theta_{\rm t}$=145.4$^{\circ}$ as derived from the most recent \cite{kallivayalil13} center of 
mass proper motion measurement. We have also initially fixed the inclination angle of our disk component at $i$=5$^{\circ}$, which is comparable to that 
determined by \cite{subramanian12} and \cite{haschke12} from the distribution of the red clump and RR Lyrae members of the Cloud and adopted the optical determination 
of the center. In lieu of the results of the previous section, we have also assumed a somewhat smaller intrinsic line-of-sight velocity dispersion of $\sigma_{v_{los}}$=22.3 kms$^{-1}$.
This model was matched to the observations by locating the global minimum of a $\chi^{2}$ goodness-of-fit statistic, allowing the parameters $\Theta$, $R_{0}$, $\eta$,
$V_{0}$, $\frac{\it di}{\it dt}$ and $v_{sys}$ (following the notation of \citeauthor{vandermarel02}, respectively, the position-angle of the line of nodes, east of 
north, the disk velocity scaling radius, disk velocity scaling index, maximum disk velocity, the rate of change of the disk inclination angle and the systemic velocity)
to vary freely. Our best fit model representation of the data, with parameters $\Theta$=127$\pm$9$^{\circ}$, $R_{0}$=0.04$\pm$0.04kpc, $\eta$=3.69$\pm$0.9, 
$V_{0}$=91$\pm$11kms$^{-1}$, $\frac{\it di}{\it dt}$=0.48$\pm$0.19mas~yr$^{-1}$ and $v_{sys}$=148.3$\pm$0.8kms$^{-1}$, has a $\chi$$^{2}$=778.2 for 760 degrees of freedom 
(the uncertainties were determined using a bootstrap with random replacement approach). The corresponding model without a disk component has a $\chi$$^{2}$=824.9 
for 765 degrees of freedom.

Considering the distribution of the F-statistic for $\nu_{1}$=5 and $\nu_{2}$=760, we find the addition of the disk component to the kinematic model provides a significant 
improvement to our reproduction of the observations. However, empirical studies of low luminosity, late-type systems have highlighted that their rotation curves attain 
lower circular velocities than those of more massive galaxies \citep[e.g.][]{rubin85,broeils92,swaters09}. With an integrated apparent B magnitude of 2.58$\pm$0.07 mag. 
\citep{bothun88}, a distance of 60kpc and $B$-$R$=1 mag. \citep{swaters99}, we estimate the absolute $R$ magnitude of the SMC to be only $M_{R}$$\approx$-17.3 mag. and 
anticipate the circular velocity of the cold HI disk to reach only $V_{c}$$\approx$60-70kms$^{-1}$ several disk scale lengths from the center. While these latter values are 
comparable to the velocities determined at the limits of the HI study of \cite{stanimirovic04}, they are substantially less than those delineated by the rotation curve
of our virtually face on disk model, even before any allowance is made for the asymmetric drift velocity of the red-giant population. Moreover, our estimate of $V_{0}$ is 
even larger if we adopt instead the HI center as the kinematic center.

We have performed a rough assessment of the asymmetric drift by noting that for an inclination angle of $i$=5$^{\circ}$, the observed velocity dispersion is effectively 
$\sigma_{z}$, the vertical component in the disk. Adopting the disk geometry from the fit above, we have determined the radial gradient of the spatial density of stars in
the SMC ($v$) by modelling the azimuthally averaged brightness profile of the galaxy as measured from 3.36$\mu$m WISE imaging \citep{wright10}. We used a function of the 
form $v=v_{0}\exp(\frac{-R}{R_{exp}})$, where the exponential scale length, $R_{exp}$, is found to have a value of 0.76$\pm$0.08kpc. Subsequently, assuming that in the putative 
SMC disk $\sigma_{z}\approx 0.7\sigma_{R} \approx\sigma_{\phi}$ (the latter two variables represent the radial and azimuthal components of the disk velocity dispersion, 
respectively), which \cite{hunter05} argues to be appropriate at least for spiral galaxies (e.g. NGC\,488, the Milky Way), we have followed \cite{hinz01} to estimate the 
asymmetric drift velocity ranges from 22kms$^{-1}$ at 0.75kpc to 55kms$^{-1}$ at 2.75kpc. In light of these sizeable corrections, it appears quite unlikely that the velocity 
signature we observe in the red-giant stars can be attributed to a disk that is viewed at a low inclination angle.  This conclusion is independent of the parameterisation
of the disk rotation curve.

Needless to say, we can obtain plausible values for the circular velocity of the disk if the inclination angle is assumed to be larger. For example, with $i$$
\approx$60$^{\circ}$, which is comparable to inclination estimates derived from the younger stellar populations in the Cloud \citep[e.g.][]{haschke12,laney86}, the 
best fitting disk model has $V_{0}$$\approx$20kms$^{-1}$ (with $\Theta$=122$^{\circ}$, $R_{0}$$\approx$8.7kpc, $\eta$$\approx$0.01, $\frac{\it di}{\it dt}$$\approx$0.5 mas~yr$^{-1}$ and 
$v_{sys}$$\approx$148kms$^{-1}$).  The initial rise of this parameterised rotation curve is rather abrupt and seemingly unphysical. However, the asymmetric drift velocity of 
the RGB star population beyond the inner few 100pc is substantial and renders the raw rotation curve, which we are modelling, effectively flat at all larger disk radii.
To explore alternative inclination angles further, we have also probed the red-giant line-of-sight velocities with a tilted ring model \citep[][]{begemann87}.
For this part of the investigation we have adopted six radius intervals with width 0.5kpc from 0.0--3.0kpc and have fixed the inclination angle at $i$=25$^{\circ}$, which is 
similar to that estimated in the pioneering study of the SMC by \cite{devaucouleurs55} and intermediate to the values discussed above. We have also anchored the kinematic 
center and the rate of change of the inclination angle at the values preferred from our analysis above. The line-of-nodes position-angle and the velocity of each ring were
permitted to vary freely in the fitting process. The results, both raw and after accounting for the generally substantial asymmetric drift velocity, are shown in figure~\ref{tring}, where for
the latter, we have taken $\sigma_{v_{los}}\approx\sigma_{\phi}$ \citep[e.g.][]{hunter05}.

At this assumed inclination angle, the inferred rotation curve rises relatively swiftly to 20kms$^{-1}$ and continues to increase gradually to 
approximately 50kms$^{-1}$ at a radial distance of 2-2.5kpc. Within the significant uncertainties, it is comparable to the empirical data for other low 
luminosity, gas-rich systems which almost invariably show a slow, steady rise in circular velocity with increasing galacto-centric distance that generally 
continues to the outermost measurement point \citep[e.g. UGC\,7603, UGC\,7971, UGC\,8490][]{swaters09}. It is also similar to the HI disk velocities
measured for the dwarf irregular NGC\,6822, which rise to $V_{c}\approx$50kms$^{-1}$ around 3kpc from the center \citep{weldrake03}. This dark matter dominated
galaxy is similar in luminosity ($M_{B}$$\approx$-15.8) to the SMC but has a much lower velocity dispersion. The line-of-sight velocity dispersion of the
SMC red-giants is effectively constant with galacto-centric distance, with no evidence for a decrease towards larger radii as anticipated for a disk 
embedded in a dark halo ($\sigma_{v}$$\propto$R$^{-1}$). A mild positive correlation between radial distance and the velocity dispersion of the carbon 
stars several kpc out in the LMC has been attributed to disk flaring \citep{alves00}. 

The impact of systemic rotation with a magnitude of $V$$\approx$25kms$^{-1}$ at an inclination angle of 25$^{\circ}$ on estimates of the SMC centre-of-mass tangential 
velocity is only a few percent and within the uncertainties that propagate from the HST proper motion measurements
of \cite[e.g. $\mu_{\rm disk}$$\simless$0.04 mas~yr$^{-1}$][]{kallivayalil13}. Given the locations of their five proper motion measurements, the details of this minor
 effect are heavily dependent on the position of the disk rotation center. For example, if this corresponds to the optical center, then the contribution of the 
systemic motion is predominantly in a SE direction, so the current centre-of-mass tangential velocity determination will be slightly overestimated. Conversely, 
if the HI center represents the stellar rotation center then the disk component to the tangential motion is predominantly in a N/NW direction, resulting in a 
slight underestimate of the centre-of-mass tangential velocity. The effect of systemic rotation is even smaller if the disk is being viewed at a larger 
inclination angle e.g. $i$$\approx$60$^{\circ}$.

We note that the position-angle inferred for the disk line-of-nodes changes abruptly by around 180$^{\circ}$ at a projected galacto-centric distance of approximately 
2.5kpc. This might have been anticipated from an inspection of figure~\ref{velofield}, which reveals a large region of generally positive velocities towards the 
far WNW of our survey region \citep[higher velocities are also evident in this region in the RGB star radial velocity study of][]{depropris10}. If the red-giant 
stars here are also part of a disk structure  in the same plane, then they appear to be counter-rotating with respect to the inner 2-3kpc of the SMC. 
Alternatively, these objects might compose a disk rotating in a similar sense to the inner regions that is merely viewed at a quite different inclination angle. 
However, counter-rotation of 
gas has been observed in other irregular galaxies
such as NGC\,4448 and IC\,10 \citep[e.g.][]{hunter98}. The latter object is a slightly less luminous ($M_{B}$$\approx$-15.5) cousin of the SMC that has recently entered an 
epoch of intense star formation \citep{wilcots98}. The gas in its central regions displays a modest rotational signature, peaking at 30-35kms$^{-1}$, but at galacto-centric
distances greater than 1--2kpc there are several HI structures with velocities that run counter to expectations based on the continuation of the kinematic trends of the inner
galaxy. \cite{wilcots98} have suggested that IC\,10 may have recently accreted this material from an extended reservoir, leading to the current burst of star formation.

The accretion of a gas rich dwarf galaxy has been invoked by several investigators \citep[e.g.][]{zaritsky00,rafelski05,bekki08,subramanian12} to account for the stark differences 
in the morphologies of the SMC's young and older stellar populations and/or the substantial variations in the Cloud's star formation rate within the last few Gyr. However, as 
discussed further in Dobbie et al. (2014), recent theoretical exploration of the chemical evolution of the SMC, which treats a major gas rich accretion event, anticipates a 
fairly pronounced dip in the age/metallicity relation \citep{tsujimoto09} for which there is little support from recent empirical determinations that consider both star clusters 
and field stars \citep{piatti11, piatti12}. Additionally, studies of the LMC indicate it has experienced substantial increases in star formation activity at similar times to the
SMC \citep[e.g.][]{harris09,smecker02}, disfavouring a triggering mechanism that was specific to the evolutionary history of the latter. Considering this and given the proximity 
of the SMC to the LMC and the Galaxy, their history of tidal interaction and the results of N-body simulations within the literature, in the next section we discuss another 
explanation, which we consider more likely, for these apparently counter-rotating stars.

\begin{figure}
\label{tring}
\includegraphics[angle=0, width=\linewidth]{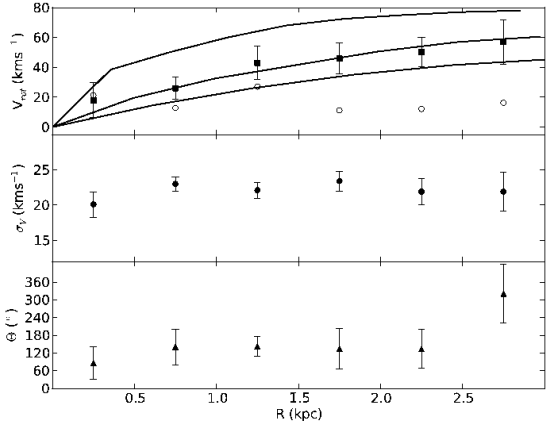}
\caption{Velocities (top, open circles), velocity dispersions (middle, filled circles) and line-of-nodes position-angles (lower, filled triangles) from 
our tilted ring analysis of the metal rich red-giant sample. Estimates of the circular velocity of the putative SMC disk, accounting for asymmetric drift,
are also shown (top, filled squares). The HI rotation curves of several dwarf galaxies e.g. UGC\,7603 (middle curve), UGC\,7971 (bottom curve), UGC\,8490 (upper curve), are 
also overplotted in the top panel.}
\label{tring}
\end{figure}

\subsection{Tidal speculation}

The larger the galacto-centric distance of a star, the more susceptible it typically is to being dislodged by the forces which arise in a galaxy-galaxy interaction 
\citep[e.g.][]{mihos94}. A two-sample Kolmogorov-Smirnov test on the normalised cumulative radial distributions for the upper and the lower metallicity quartiles of 
our red-giants \citep[constructed by adopting the semi-major axis of an ellipse, with $a$/$b$=1.5 and an origin at the center of the Cloud, on which a star lies, as 
a proxy for its galacto-centric distance, e.g.][]{piatti07,subramanian12} confirms that the spatial distributions of these sub-populations are significantly different.
The metal-poor quartile displays a considerably more extended distribution, in accord with the detection by \cite{carrera08} of a radial metallicity gradient in the 
SMC field red-giants. The kinematic structure towards the NW limits of our survey region, which is weakly evident in our C-star velocity surface and in HI surveys 
of the Cloud \citep[e.g. figure 5 of ][]{stanimirovic04}, is most apparent in the data for our metal-poor RGB star sub-sample (e.g. figures~\ref{velofieldmp}, 
\ref{velofieldmr} and \ref{rcurve}).

\cite{bekki09} predict an extension in the surface density of the intermediate/old stellar SMC populations towards the NW (their figure 6), which they attribute 
to prior interaction with the LMC and/or the Galaxy. According to the simulations of \cite{gardiner96}, a tidal tail formed by the most recent encounter with the LMC, 
extends from the main body of the galaxy, initially in a broadly south-westerly direction, before turning back towards the NE, behind the Cloud. This structure appears
to be analagous to the gaseous ``Counter-Bridge'' of the more recent modelling of \cite{diaz12}, that similarly is predicted to extend out behind the SMC. Both the 
Magellanic-Bridge and the Counter-Bridge are expected to display a positive correlation between velocity and distance \citep[e.g.][]{gardiner96}, which, at least in 
the former case, appears to have empirical support. For example, the closer of the two relatively distinctive structures identified by \cite{nidever13} towards the 
N and E of the main body, which they propose to be an intermediate-age stellar counterpart of the Magellanic Bridge, is associated with a secondary, lower velocity 
component observed in this direction \citep[e.g.][ and our figure 8]{hatzidimitriou97}.

Our velocity histograms towards the W and NW of the Cloud, fields 3D5, 3D9 and 3D11, suggest a secondary, higher velocity peak here (170-180kms$^{-1}$) but it is 
considerably less well populated than the low velocity structure on the NE side. \citeauthor{nidever13} find no compelling evidence of multiple stellar populations 
along the lines-of-sight towards their W and NW pointings, although the peak in the distance distribution of the red clump stars in their field 40S296 is at a marginally 
greater distance (approximately 71 kpc) than that in the bulk of their other samples. The neighbouring pointings 40S251 and 40S206, despite their lower overall populations,
also hint at somewhat more distant stars. Despite a lack of detailed agreement in terms of the projected spatial distribution of these structures, between observations and 
the results of the simulations of the interaction of the SMC, LMC and Galaxy, in consideration, we find tidal formation of this apparently counter-rotating, kinematical 
structure towards the NW to be preferred over a galaxy merger/accretion origin. Consequently, we propose these stars are a stellar component of the ``Counter-Bridge'' tidal 
tail.  Our finding that the higher velocity peak towards the NW is much weaker than the low velocity structure on the NE side, concurs with the predictions of a recent maximum interaction simulation of the Magellanic system. 
While models in which the Clouds merely experience a close encounter \cite[e.g.][ model 1]{diaz12,besla12} appear to show the Bridge and the Counter-Bridge have comparable stellar densities, the latter is postulated to be considerably less well populated than the former in a simulation involving a direct collision between the LMC and SMC 
around 100Myr ago \citep[][]{besla12}. A very strong interaction between the Clouds seems neccessary to account for the sizeable observed population of SMC AGB stars that have been accreted onto the LMC \citep{olsen11}.

The influence of tidal forces on the SMC could also be a plausible explanation for the moderate rate of change of the inclination-angle we infer for the putative disk
in our modelling of the line-of-sight velocities of the centrally concentrated metal-rich red-giants ($\frac{\it di}{\it dt}$$\approx$140$^{\circ}$ Gyr$^{-1}$). \cite{piatek95} 
have demonstrated using an N-body simulation of a dwarf spheroidal interacting with a larger galaxy that the effects of tidal forces can be manifest as an apparent rotation 
around the minor axis of the satellite. However, it is not clear if the viewing geometry here is conducive to this. These simulations also indicate that stars residing 
at smaller galacto-centric radii are dislodged in a galaxy-galaxy interaction since an additional factor which dictates their fates is velocity along the line-of-centers 
\citep{piatek95}. While the changing inclination of the LMC's disk has been attributed to the impact of tidal torques from the Galaxy \cite{vandermarel02}, further investigation
of the SMC kinematics indicates that the non-negligible value of $\frac{\it di}{\it dt}$ here stems from the protrusion of lower velocities towards the Wing region, lower 
velocities towards the NE of our survey area and an excess of positive stellar GRF velocities just beyond the SW end of the SMC Bar (see figure~\ref{velofieldmr}). The two zones 
of negative GRF velocities, at least, appear to tie in with features previously touted as having formed as a result of the most recent encounter with the LMC. In view of the 
kinematics and the locations of the stars towards the Wing and around the SW end of the Bar it is tempting to associate them with, respectively, the lower velocity foreground 
complex proposed by \cite{nidever13} as an intermediate-age stellar component of the Magellanic-Bridge and the structure in the W and NW of our survey area that we have advanced 
as a stellar analogue of the ``Counter-Bridge''. However, the signs of the velocities in both regions appear to be contrary to expectations from simulations of the SMCs evolution 
\citep[e.g.][]{diaz12,gardiner96}. Perhaps at least part of this discrepancy is due to an incorrect assumption in these calculations of the orientation of the SMC stellar disk. 
Nonetheless, these new observations and prior theoretical and empirical studies \citep[e.g.][]{bekki09,olsen11} paint a complex picture of an intermediate-age stellar population 
in the SMC that is subject to substantial tidal stripping. Given the expected distances between the SMC and Milky Way compared to the SMC-LMC separation over the past 3--5Gyr, the magnitude of the tidal force of the LMC on the SMC may have been more than an order of magnitude higher than the MW-SMC tidal force. It is not unreasonable to link the SMC structure and history to its orbit around the LMC rather than the Milky Way.

\section{Summary}

\begin{figure}
\label{concept}
\includegraphics[angle=0, width=\linewidth]{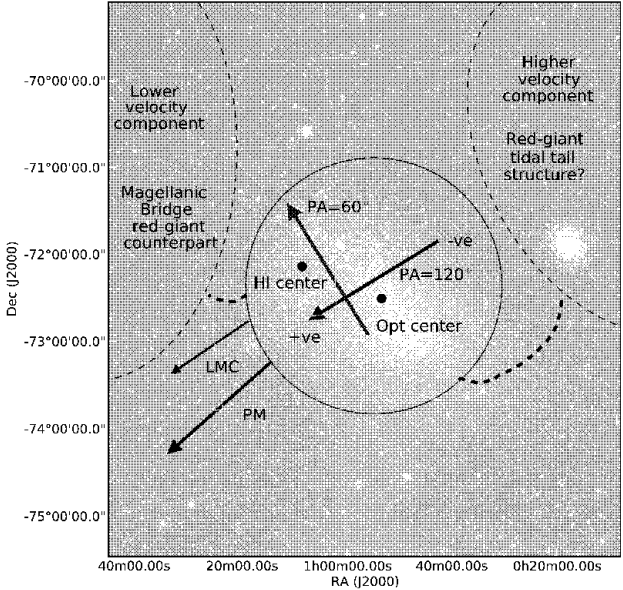}
\caption{A conceptual plot of the SMC highlighting the position-angles of the HI \citep{stanimirovic04} and the RGB star kinematic axes (this work). The locations 
of the NE (possibly the Magellanic Bridge) and the suspected NW (possibly the ``Counter-Bridge'') tidal structures are highlighted, as are the optical 
and the HI centers of the Cloud, the SMC's direction of tangential motion and direction of the LMC.}
\label{concept}
\end{figure}

We have acquired optical spectroscopy for a sample of 3037 predominantly RGB stars distributed across 37.5 deg$^{2}$ of the SMC and have 
measured their radial velocities to an accuracy of better than 5kms$^{-1}$. The main results of our analysis of these datasets, as illustrated in 
figure~\ref{concept}, are:

\begin{itemize}

\item{We have found a velocity gradient in the rest frame of the SMC which is consistent in phase with that observed in the young, massive stellar population
by \cite{evans08}. The line-of-sight velocity and velocity dispersion profiles for the most metal poor giants, the most metal rich red-giants and the F/G supergiants, 
along a PA$\approx$120$^{\circ}$, the direction of the steepest gradient, and the arrangement of the Cloud's coherent magnetic field across the Bar suggest this arises from 
the systemic rotation of disk structure.}

\item{The orientation of this putative stellar disk is discordant with the position angle of the SMC kinematic axis derived from HI observations (PA$\approx$60$^{\circ}$). 
This inconsistency is challenging considering that the F/G supergiants examined here would have had only circa 200Myr in which to dynamically separate from the HI gas. 
The source of the discrepancy may be due to degeneracies in the model formalism for linear (or nearly linear) rotation curves (see below).}

\item{An almost face-on viewing angle ($i$$\approx$5$^{\circ}$) has been favoured by recent studies of the intermediate and old stellar populations of the SMC. Analyses 
of the line-of-sight velocity of the most metal-rich SMC red-giant stars, following the disk formalism of \citet{vandermarel02} and with a tilted ring model, show that 
for such low inclinations the inferred circular velocities (V$_{c}\simgreat$90kms$^{-1}$) are much larger than those typically observed in a galaxy of the SMC's luminosity 
(V$_{c}\approx$60--70kms$^{-1}$). It is thus unlikely that the velocity patterns seen in these stars are attributable to a disk viewed nearly face-on.} 

\item{Assuming a larger inclination angle (e.g. $i\approx 25^{\circ}$), comparable to that determined by \cite{devaucouleurs55}, we can infer a rotation curve for an 
SMC disk that, within the uncertainties of our analysis, is accordant with those of similarly luminous galaxies such as NGC\,6822. If a disk with a kinematic major-axis along 
a PA$\approx$120$^{\circ}$ is the correct interpretation of the observed velocity patterns within the central regions of our survey area then some revision to simulations of the
evolution of the Cloud may be required as these are generally tuned to the disk plane lying in a NE-SW direction at the present epoch. However, the neglect of this systemic
 rotation has only a minor impact on estimates of the Cloud's centre-of-mass tangential motion.}

\item{Considering the proximity of the SMC to the LMC and the Galaxy, their history of tidal interaction and the different spatial distributions of the young and old stars
within the Cloud, we associate a kinematical structure towards the far north-west of our survey area, that is particularly evident in the velocities of our most metal 
poor red-giants, with the tidal ``Counter-Bridge'' tail predicted by N-body and hydrodynamical simulations of the SMCs most recent interaction with the LMC.}

\item{The moderate rate of change of the disk inclination angle, $\frac{\it di}{\it dt}$$\approx$ 140$^{\circ}$ Gyr$^{-1}$, inferred by our modelling, is also likely attributable to 
the influence of tidal forces on the SMC. The protrusion of lower velocities towards the Wing region and an excess of positive stellar GRF velocities just beyond the SW end of 
the SMC Bar, observed in kinematics of the younger, metal-rich RGB stars, are probably associated with a lower velocity foreground complex proposed by \cite{nidever13} as an 
intermediate-age stellar component of the Magellanic-Bridge and the structure in the W and NW of our survey area that we advance as a stellar analogue of the ``Counter-Bridge'', 
respectively.}

\item{The disagreement between disk parameters derived from HI and stellar velocities opens the
question of how to appropriately model a system in which multiple processes contribute similar
amounts to the observed velocities. For example, the inferred rate of change of inclination of the HI
disk of about 400$^{\circ}$ Gyr$^{-1}$ (Indu \& Subramaniam 2014, in prep) is much larger than that for the 
stellar disk and represents a radial velocity gradient of approximately 5.5 kms$^{-1}$kpc$^{-1}$ 
along the major axis of the disk. This is of the same order as the apparent velocity gradient 
induced by the transverse motion of the centre of mass, and not much smaller than the disk rotation velocities 
we derive. The $\frac{\it di}{\it dt}$ and $V_{c}$
 contributions are defined to be perpendicular to each other, and so model algorithms
are likely to trade one against the other when the rotation curve is linear. In the SMC case, we also
have to consider that the proper motion is directed nearly parallel to the stellar disk position
angle. This degeneracy is likely to be broken only when significant samples of stars beyond the radius at which
the rotation curve flattens out are measured, and the data considered in the light of the origins of any torques that 
could significantly alter the disk angular momentum.}

\item{A more spatially extensive radial velocity survey of the young stellar population, including sources located north 
of the $\delta$=-72$^{\circ}$ limit of the \citeauthor{evans08} work, would help address questions relating to the putative 
SMC disk structure we have tentatively identified from the kinematics of the youngest red-giant stars in our sample. A 
spectroscopic survey of the SMC's red-giant population at larger projected galactocentric distances than included by our 
survey area would also be useful to search for a continuation (or otherwise) of kinematic trends identified here and would 
provide greater insight on the existence and extent of the tidal structures.} 

\end{itemize}

\section*{Acknowledgments}
This publication makes use of data products from the Two Micron All Sky Survey, which is a joint project of the 
University of Massachusetts and the Infrared Processing and Analysis Center/California Institute of Technology, 
funded by the National Aeronautics and Space Administration and the National Science Foundation. This research was
supported under the Australian Goverment's Australia-India Strategic Research funding scheme (reference number 
ST040124).

%
\bibliographystyle{mn2e}
\bibliography{mnemonic,therefs}

\bsp

\label{lastpage}

\end{document}